\documentclass[a4paper,fleqn,usenatbib,useAMS]{mnras}

\usepackage{graphicx}	% Including figure files
\usepackage{amsmath}	% Advanced maths commands
\usepackage{amssymb}	% Extra maths symbols
\usepackage{multicol}        % Multi-column entries in tables
\usepackage{pdflscape}	% Landscape pages
\usepackage{amstext}
\usepackage{lineno}
\usepackage{graphicx}
\usepackage{soul}

\title[$L_{*}$ Ellipticals]{\textit{Witnessing the star-formation quenching in $L_{*}$ ellipticals}}

\author[Dhiwar et al.]{Suraj Dhiwar,$^{1,2}$\thanks{Contact e-mail: suraj@iucaa.in \href{}{}},
Kanak Saha,$^{1}$
Avishai Dekel,$^{3,4}$
Abhishek Paswan,$^{1,5,6}$
Divya Pandey,$^{7}$
\newauthor % used to write full names.
Arianna Cortesi,$^{8,9}$
and Mahadev Pandge$^{10}$
\\
$^{1}$Inter University Center for Astronomy and Astrophysics, Ganeshkhind, Pune 411007, India\\
$^{2}$Department of Physics, Savitribai Phule Pune University, Pune 411007, India\\
$^{3}$Racah Institute of Physics, The Hebrew University, Jerusalem 91904 Israel\\
$^{4}$SCIPP, University of California, Santa Cruz, CA 95064, USA\\
$^{5}$Indian Institute of Astrophysics (IIA), Koramangala, Bengaluru 560 034, India.\\
$^{6}$ Deapartment of Physics, University of Allahabad, Prayagraj 211002, India\\ 
$^{7}$Department of Physics and Astronomy, National Institute of Technology, Rourkela, Odisha 769 008, India\\
$^{8}$Instituto de Astronomia, Geof\'{i}sica e Ci\^encias Atmosf\'ericas (IAG), Universidade de S\~{a}o Paulo (USP),\\
R. do Mat\~{a}o 1226, 05508-090, S\~{a}o Paulo, Brazil\\
$^{9}$Observatorio de Valongo, Universidade Federal do Rio de Janeiro
, Ladeira do Pedro Ant\^onio, 43, Centro,\\ Rio de Janeiro - RJ 20080-090, Brazil\\
$^{10}$Dayanand Science College, Latur, Maharashtra, 413512, India
}

\date{Last updated ; in original form }

\pubyear{2022}
\begin{document}
\label{firstpage}
\pagerange{\pageref{firstpage}--\pageref{lastpage}}
\maketitle

\begin{abstract}
We study the evolution of $L_{*}$ elliptical galaxies in the color-magnitude diagram in terms of their star-formation history and environment, in an attempt to learn about their quenching process. We have visually extracted 1109 $L_{*}$ galaxies from a sample of 36500 galaxies that were spectroscopically selected from Stripe82 of the Sloan Digital Sky Survey. From this sample we have selected 51 ellipticals based on their surface-brightness profile being well-fitted by a single S$\acute{e}$rsic profile with S$\acute{e}$rsic indices $3<n<6$. Our sample consists of 12 blue-cloud $L_{*}$ ellipticals (BLE), 11 green-valley $L_{*}$ ellipticals (GLE), and 28 red-sequence $L_{*}$ ellipticals (RLE). We find that most of the RLEs and GLEs have been quenched only recently, or are still forming stars, based on their [{O\sc{iii}}] and H$\alpha$ emission, while the BLEs are forming stars vigorously. The star-formation in BLEs is found to be extended over the galaxy and not confined to their central region. In about 40\% of the $L_{*}$ ellipticals (ten BLEs, four GLEs and five RLEs), star-formation quenching seems to have started only recently, based on the lower [{O\sc{iii}}] emission compared to the [{O\sc{ii}}] and H$\alpha$ emission, at a given metallicity. We also find that the galaxy color is correlated with the cosmic-web environment, with the BLEs tending to reside in lower-density regions, the RLEs preferring denser, clustered regions, and the GLEs found in either. One possible scenario is that as the star-forming ellipticals migrate into the clusters, their star formation is suffocated by the hot intra-cluster medium.

\end{abstract}

\begin{keywords}
galaxies: photometry --- galaxies: evolution --- galaxies: star formation --- galaxies: structure ---galaxies: elliptical and lenticular, cD
\end{keywords}

%%%%%%%%%%%%%%%%% BODY OF PAPER %%%%%%%%%%%%%%%%%%

% The MNRAS class isn't designed to include a table of contents, but for this document one is useful.
% I therefore have to do some kludging to make it work without masses of blank space.
%\begingroup
%\let\clearpage\relax
%\tableofcontents
%\endgroup
%\newpage

\section{Introduction}
\label{sec:intro}

The broad-band optical colours of galaxies in the local universe show prominent bimodality,  dividing galaxies into two broad categories: the red sequence and the blue cloud. The red sequence is composed of mainly the ellipticals, the lenticulars, and bulge-dominated disk galaxies with old stellar populations, with little or no star-formation, commonly found in a dense environment. While the blue cloud is composed of star-forming galaxies with a younger stellar population inhabiting the low-density environment. This colour bimodality in galaxies \citep*{Takamiyaetal1995} has been observed at low redshift \citep{Stratevaetal2001, Baldryetal2004} as well as at redshift $z \sim 1 - 2$ \citep{Baloghetal1998, Willmeretal2006, Brammeretal2009, Muzzinetal2013}. Not only the stellar mass has increased over the past ten billion years, but the number density of the red sequence population has also enhanced, especially above a characteristic stellar-mass limit $\sim$ $3 \times 10^{10}$~M$_{\odot}$  \citep{Kauffmanetal2003, Belletal2004, Blanton2006, Bundyetal2006, Faberetal2007, Mortlocketal2011, Ilbertetal2013, Moustakasetal2013, Sachdevaetal2019}. It is likely that a fraction of galaxies from the blue cloud have emigrated to the red sequence by suppressing their star-formation activity known as quenching and possibly accompanied by a morphological transformation \citep{Martigetal2009, Mishraetal2019}. To date, several physical processes are being put forward to explain the onset of star-formation quenching such as stripping of cold gas in cluster medium \citep{Gunn-Gott1972, Abadietal1999, Bosellietal2019}, cutting off the supply of cold gas, called strangulation \citep{PengMaiolinoCochrane2015}, halo quenching or quenching due to compaction \citep{DekelBirnboim2006,Wooetal2015} and thereby growth of the red sequence. Nevertheless, it remains a difficult task to constrain which physical processes contribute to what extent. Subtle effects of the environment add a layer of complication to the list.

\par
In order to simplify the problem of understanding quenching and what drives it, we focus only on the elliptical galaxies having smooth light profiles and relatively simpler morphology compared to disk galaxies. This allows us to keep morphological quenching aside from our discussion. Massive ellipticals display remarkable uniformity in their physical properties such as light profiles, broadband colors, stellar population, metallicity gradients, the ratio of rotational velocity to velocity dispersion. Based on observational evidence of the variations of the color-magnitude diagram (CMD), size-mass relation, and fundamental-plane relation with redshift, it is thought that bright ellipticals formed the bulk of their stellar mass as early as redshift $z \sim 2$. The traditional view is that these bright ellipticals have evolved passively since then \citep{EggenLynden-BellSandage1962, TinsleyGunn1976, DjorgovskiDavis1987, Bender1997, Matteucci1997}. Although a significant fraction of the stellar mass in bright ellipticals does form at high redshift, in recent decades, studies have shown that some elliptical galaxies retain a signature of star formation at late epochs \citep{Zabludoffetal96, Quinteroetal2004, Gotoetal2003}. Ever since the discovery of blue ellipticals \citep{Stratevaetal2001, Fukugitaetal2004}, the simplified notion about ellipticals has changed. Subsequent studies have shown that these blue ellipticals have, in general, low stellar masses \citep[e.g.][]{2009MNRAS.393.1324B, 1988A&A...204...10K,1997A&AS..124..405D, 1997MNRAS.288...78T}; occupy the blue cloud on the CMD, and exhibit ongoing star formation \citep{1999MNRAS.304..319F,Kavirajetal2007, 2009MNRAS.393.1324B, 2010gama.conf..195A, 2010A&A...515A...3H, 2013MNRAS.432..359T}. 

\par
Apparently, the host galaxy stellar mass seems to play a key role in determining the location of the ellipticals or even the redward migration on the CMD. Galaxies with stellar mass $> 6.3 \times 10^{10}$~M$_{\odot}$ in the local universe are, generally, thought to be quenched by AGN feedback \citep{Hopkinsetal2005, Kavirajetal2007, Fabian2012, Cimattietal2013, Ciconeetal2014, ForsterSchreiberetal2014,paswan2016,Kalinovaetal2021}, although there are alternatives to AGN feedback to quench star formation activity in galaxies \citep*{Khochfar-Silk2006, Naab-Khochfar-Burkert2006, NaabJohanssonOstriker2009}. Additionally, these massive galaxies could be subject to the mass quenching \citep{Pengetal2010} or shock-heating of the inflating cold gas at the virial radius of the surrounding dark matter halo \citep{DekelBirnboim2006}, which cuts off the cold gas supply fueling the star-formation. In low-mass galaxies, i.e., below $10^{10}$~M$_{\odot}$, supernovae or stellar feedback is thought to drive quenching \citep{Kavirajetal2007}. This brings us to the key question to be addressed in this work - how does the star-formation quenching take place in the intermediate-mass ($3 \times 10^{10}$~M$_{\odot} \sim $ stellar mass of an $L_{*}$ galaxy) galaxies and how to identify which of them is currently going through the quenching process.

In this work, we consider a set of 51 visually classified ellipticals from a sample of 1109 $L_{*}$ galaxies from the SDSS Stripe82 survey \citep{2009ApJS..182..543A} with a narrow mass range around $~3\times 10^{10}$M$_{\odot}$, having spectroscopic redshift (z $\sim$ 0. - 0.3) and optical colors varying from g-r$\simeq 0.2$ (bluest) to $\simeq 1.0$ (reddest). Our sample of $L_{*}$ ellipticals consist of ellipticals in the blue cloud, green valley, and red-sequence and is ideally suited (or simplified as we can remove those mechanisms applicable to spirals) to address questions like how blue or green valley ellipticals are quenched; which quenching mechanisms are likely to be in action? The availability of SDSS fibre spectra is a plus to identify which of these ellipticals are currently in the process of quenching. As the sample of $L_{*}$ ellipticals are spread over a diverse environment on the cosmic web, it naturally allows us to study the role played by the environment on the star-formation quenching. 

Throughout this work, all distance dependent quantities assume $H_{0}$ = 70 km s$^{-1}$ Mpc$^{-1}$, $\Omega_{m}$ = 0.3, and $\Omega_{\Lambda}$ = 0.7 following $\Lambda$CDM cosmology.

%%%%%%%%%%%%%%%%%%%%%%%%%%%%%%%%%%%%%%%%%%%%%%%%%%%%%%%%%%%%%%%%%%%%%%%%%%%%%

\begin{figure*}
 \includegraphics[width=\columnwidth]{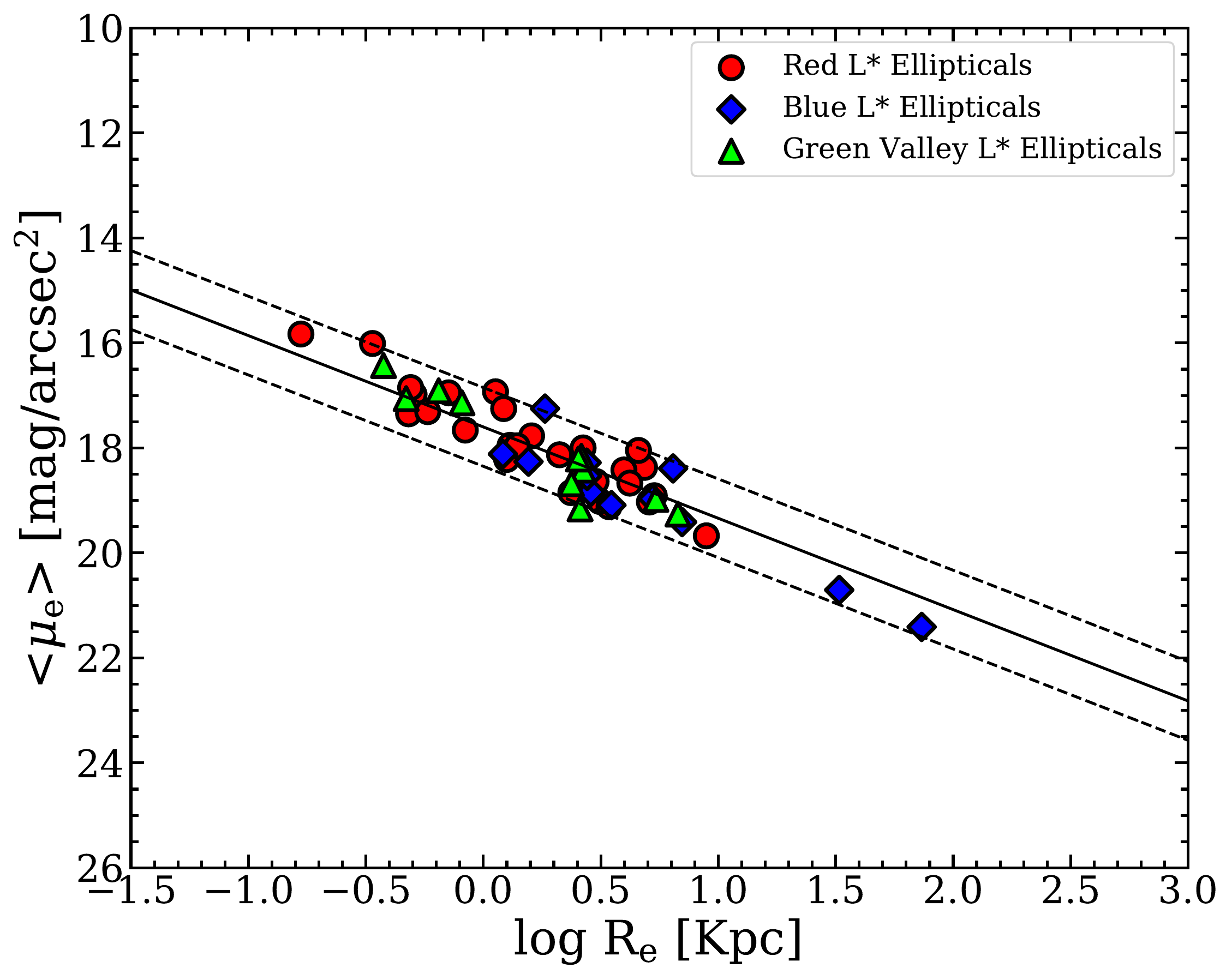}
 \includegraphics[width=\columnwidth]{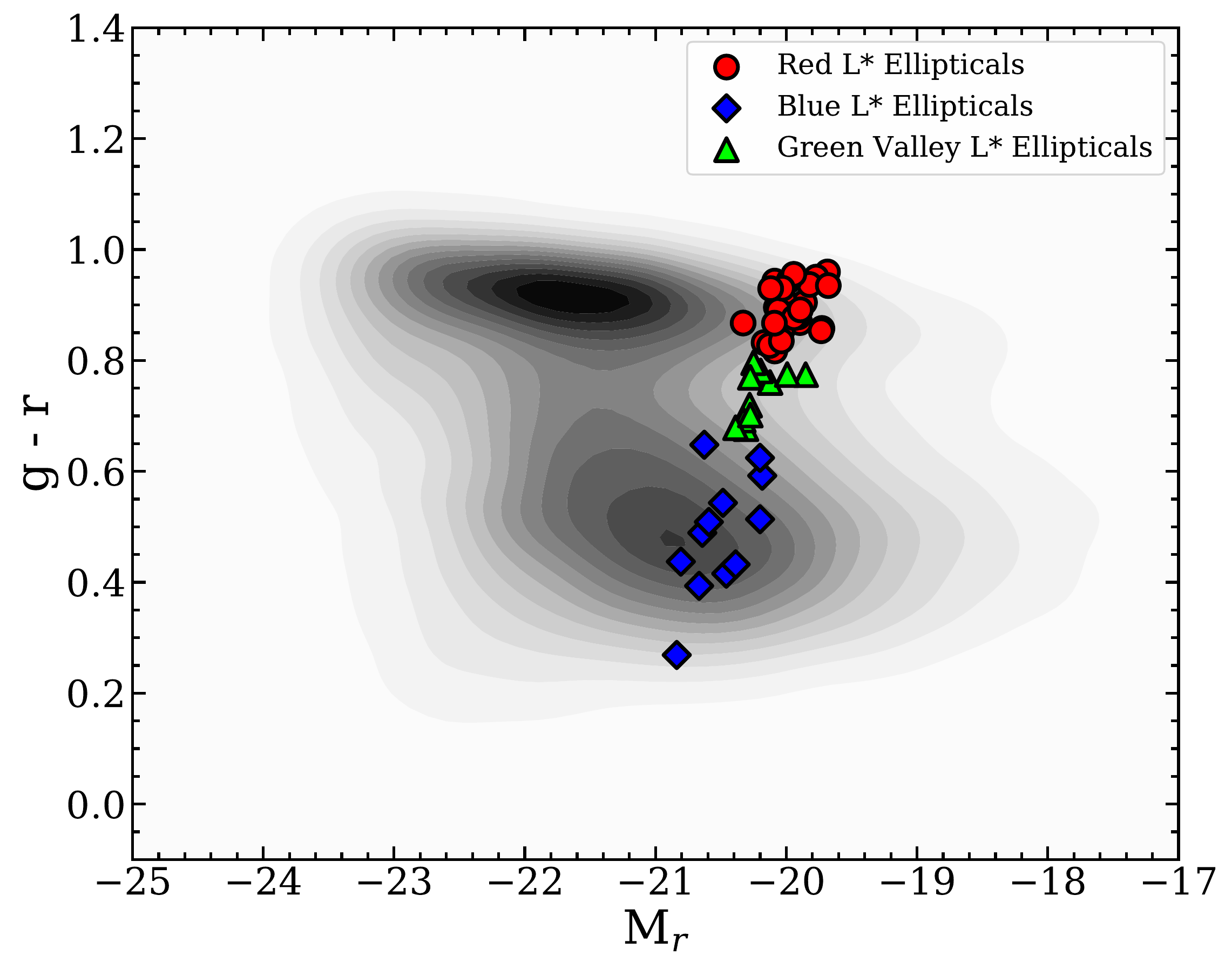}
 \caption{Left: Kormendy plot - Red $L_{*}$ ellipticals are represented by filled circles green valley $L_{*}$ ellipticals by triangles and diamonds are blue $L_{*}$ ellipticals. Solid line is from \citep{2009MNRAS.393.1531G} while the dashed lines are 3$\sigma$ scatter.
 Right: Color Magnitude Diagram- Contours are for the galaxies in Stripe 82 from SDSS while the red circles represents red $L_{*}$ ellipticals, green triangles are green valley $L_{*}$ ellipticals and blue diamonds are blue $L_{*}$ ellipticals.}
 \label{fig:cmd}
\end{figure*}

%%%%%%%%%%%%%%%%%%%%%%%%%%%%%%%%%%%%%%%%%%%%%%%%%%%%%%%%%%%%%%%%%%%%%%%%%%%%%

\section{Data}
\label{sec:data} 
We use the Stripe 82 co-added imaging data from \citep{ 2016MNRAS.456.1359F}. Stripe 82 is 275 deg$^2$ region in the equatorial plane with RA $-$50 $<$ 60 $^{\circ}$ Dec $-$1.25 $<$ 1.25 $^{\circ}$. It has been repeatedly imaged by SDSS to reach fainter magnitudes \citep{2009ApJS..182..543A}. While several other stacked data sets are also available \citep[e.g.][]{2014ApJ...794..120A, 2014ApJS..213...12J}, \citet{ 2016MNRAS.456.1359F} uses a non-aggressive sky subtraction strategy preserving the low surface brightness structures. The co-added images are 1.7 - 2.0 mag deeper than the single-epoch SDSS images and median seeing of 1.0 - 1.3 arcsec depending on the band. Based on 50$\%$ completeness  limits for point sources, the co-adds reach 24.2, 25.2, 24.7, 24.3, 23.0 magnitudes in u,g,r,i,z bands, respectively. 

%%%%%%%%%%%%%%%%%%%%%%%%%%%%%%%%%%%%%%%%%%%%%%%%%%%%%%%%%%%%%%%%%%%%%%

\subsection{Star/galaxy separation and photometry}
\label{sec:methods}

We run SExtractor \citep{1996A&AS..117..393B} on the stacked images for the detection of sources and the measurement of their photometric and structural parameters in all five SDSS bands.  Each band contains 1095 frames, their corresponding point spread function (PSF) images and weight images. Local background was estimated in a box size of 64 pixels adopting a detection threshold of 0.5$\sigma$ above the background. The 0.5$\sigma$ detection threshold was adopted in order to detect faint sources and construct the segmentation maps, which were later used in masking. We measured the central coordinates, Petrosian radii, Petrosian magnitudes, semi-major axis and CLASS STAR parameter using SExtractor. For CLASS STAR parameter determination, seeing Full-Width at Half Maximum (FWHM) were derived from the PSF images. Subsequently, 5,89,223 objects were detected combining extended and point sources. For star/galaxy classification, we utilize the CLASS STAR parameter in SExtractor and $\Delta_{sg}$ defined by \citep{2010MNRAS.404...86B} as\\

CLASS$\_$STAR $<$ 0.1 and $\Delta_{sg}$ $>$ 1\\

\noindent where $\Delta_{sg}$ = r$_{psf}$ - r$_{model}$, r$_{psf}$ is r band psf magnitude and r$_{model}$ is r band model magnitude

Similar procedure was followed for object detection and photometry in the UK Infrared Deep Sky Survey (UKIDSS) J, H, K bands. A total of 1,74,475 galaxies are detected after matching SDSS (u,g,r,i,z) and UKIDSS (J,H,K) band photometry. On procuring spectroscopic redshifts from SDSS spectroscopic Data Release 12 (DR12), our galaxy sample is downsized to 36,581 having spec-z. Since K-corrections are not reliable above redshift z $\sim$ 0.5 \citep{2011MNRAS.413.1395O}, we consider galaxies below z $\sim$ 0.5, which further reduces the sample size to 30,574 galaxies. In the rest of the paper, we refer to this as our final sample. Although the final sample is obtained after matching SDSS and UKIDSS photometry, we do not use UKIDSS bands in this analysis and will be published with the entire $L_{*}$ galaxy catalog in a subsequent paper.

Using standard cosmological parameters and the luminosity distance ($D_{L}$) estimated from the spectroscopic redshift, we obtain the absolute magnitudes for our sample of galaxies using the following relation:

\begin{equation}
M_{\lambda} = m - 5(\log(D_L)-1) - K_{\lambda} - A_{\lambda},
\end{equation}

\noindent where m is the Petrosian magnitude of the galaxies from SExtractor, K$_{\lambda}$ and A$_{\lambda}$ are the K-correction \citep[adapting a linear dependence of k-correction on z and (g-r) color)]{2011MNRAS.413.1395O} and Galactic dust extinction \citep{2011ApJ...737..103S} correction, respectively. These corrected absolute magnitudes were used to obtain the integrated colors (g-r, u-r and r-z) of galaxies. 

%%%%%%%%%%%%%%%%%%%%%%%%%%%%%%%%%%%%%%%%%%%%%%%%%%%%%%%%%%%%%%%%%%%%%%%%%%%%%

\subsection{Stellar Mass estimate}

The stellar mass, being one of the most fundamental properties of a galaxy, governs the evolution and formation processes. Hence, estimating the stellar mass of a galaxy accurately is crucial when studying galaxy evolution. Several methods exist to determine galaxy masses. Most methods consist of measuring M$_*$ from mass to light ratio (M/L) of the stellar populations. The M/L can be obtained from the fitting of stellar population synthesis (SPS) models to the Spectral Energy Distribution \citep[SED;][]{Walcheretal2011, Conroy2013, Courteauetal2014} of a galaxy. However, the color based M/L estimates are widely used, and several calibrations exist in the literature \citep[e.g.][]{BelldeJong2001, Belletal2003, GallazziBell2009, Tayloretal2011, McGaughSchombert2014}. These calibrations take into account the underlying stellar populations, dust attenuation and chemical evolution. Stellar masses for our sample are calculated using the r-band luminosity and a mass-to-light ratio calculated following the Bell-de Jong relation \citep{BelldeJong2001}: 

\begin{equation}
  log(\gamma = M/L) = 1.097\times(g-r)-0.306  
\end{equation}

\noindent where $\gamma$ is the mass-to-light ratio and $g-r$ is the optical color. The calibrations of \citet{BelldeJong2001} which uses diet Salpeter IMF, SPS, galaxy evolution and dust, yields a wavelength dependent error of $\sim$ 0.1-0.2 dex in M$_*$/L and increases to 0.3 - 0.6 dex if NIR colors are used. (Note that in our sample, the M/L estimates are based on optical colors).
 It has been demonstrated that color based M/L do not deviate much from that of the SED estimates \citep{RoedigerCourteau2014}. Using a mock galaxy sample, they show that the masses based on optical colors and SED fitting agree within a scatter of 0.2 dex and reproduced similar results when applied to observed galaxies with M$_* \sim 10^{8}-10^{11} M_{\odot}$.

%%%%%%%%%%%%%%%%%%%%%%%%%%%%%%%%%%%%%%%%%%%%%%%%%%%%%%%%%%%%%%%%%%%%%%%%%%%%%%%%%%%%%%%%%%%%%%%%%%

\begin{figure*}
 \includegraphics[width=0.9\textwidth]{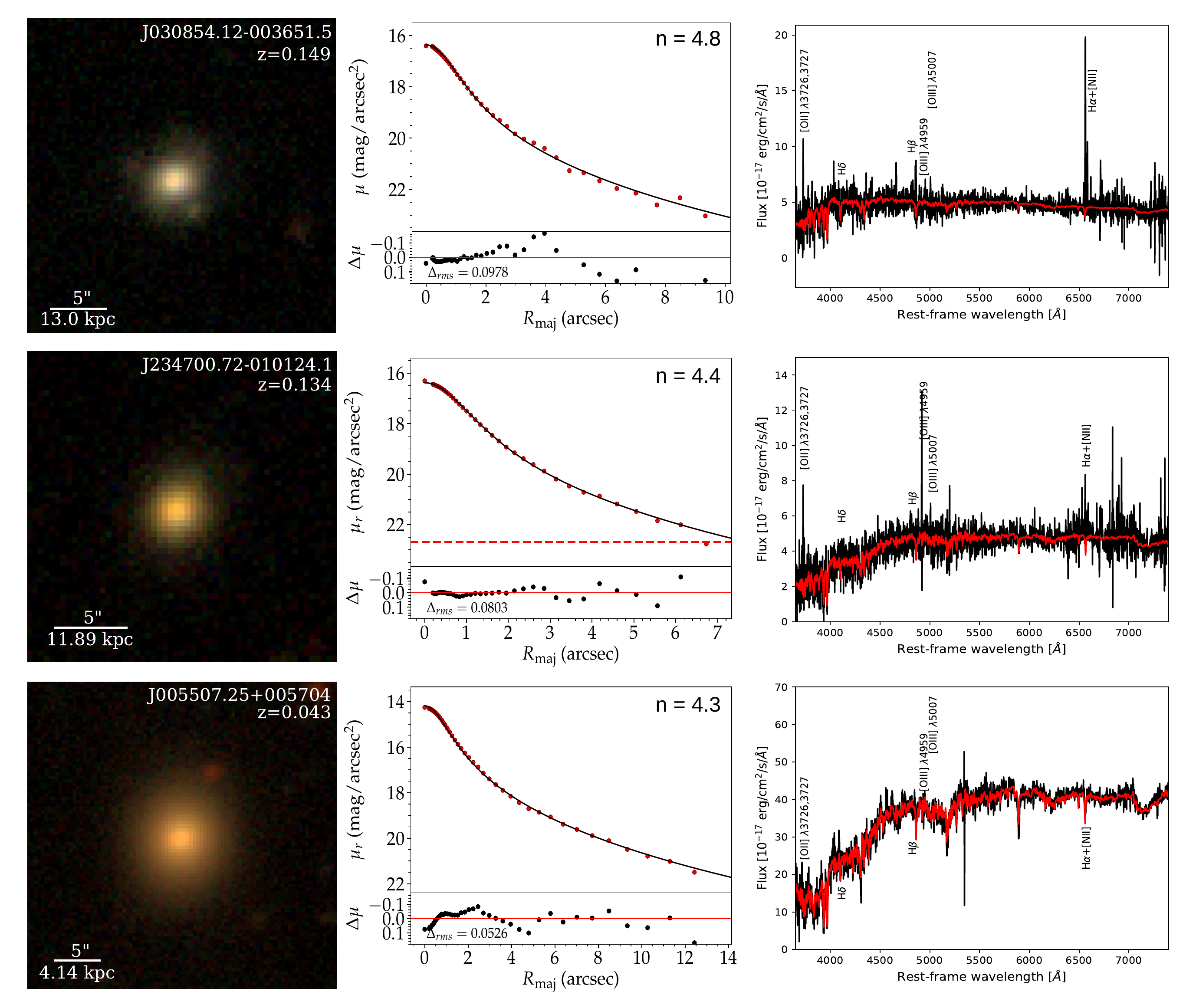}
 \caption{The gri color image, corresponding surface brightness profiles (red filled circles), a S$\acute{e}$rsic fit to the profile (black solid curve) and respective SDSS observed spectra (black line) fitted with stellar continuum and absorption features (red line) of a blue $L_{*}$ elliptical (top), green valley $L_{*}$ elliptical (middle) and red $L_{*}$ elliptical (bottom) from the sample.}
 \label{fig:stamps}
\end{figure*}

%%%%%%%%%%%%%%%%%%%%%%%%%%%%%%%%%%%%%%%%%%%%%%%%%%%%%%%%%%%%%%%%%%%%%%%%%%%%%
\begin{table*}
 
\begin{center}
%\caption{Properties of $L_{*}$ galaxies}
\begin{tabular}{lcccccccc}
\hline \hline
SDSS Obj ID & z & m$_r$  & n$_r$ &  n$_g$ &  R$_e$ & E(B-V) & Mass & SFR \\
 &   & mag &  & & kpc & & $\times$10$^{10}$M$_{\odot}$ & M$_{\odot}$/yr \\
(1) & (2)  & (3) & (4) & (5) & (6) & (7) & (8) & (9)\\ \hline
Blue $L_{*}$ Ellipticals \\
\hline
J225502.14-000504.8 & 0.184 & 19.73 & 4.8 & 3.6 & 6.46 & 1.38 & 3.38 & 17.1300 \\ 
J000919.89-010229.6 & 0.081 & 17.14 & 3.4 & 3.5 & 1.30 & 0.59 & 3.40 & 7.3840 \\ 
J215022.22+001017.7 & 0.206 & 19.88 & 4.3 & 4.2 & 1.58 & 0.51 & 3.17 & 7.2015 \\ 
J224259.05+004624.8 & 0.160 & 18.89 & 3.7 & 3.2 & 1.55 & 0.28 & 3.45 & 5.3896 \\ 
J211651.79-010141.4 & 0.132 & 18.20 & 3.6 & 3.1 & 2.24 & 0.39 & 3.04 & 3.9240 \\ 
J025652.34+004555.3 & 0.143 & 18.67 & 4.9 & 3.9 & 1.56 & 0.38 & 3.20 & 3.9065 \\ 
J025302.59-011305.1 & 0.233 & 19.60 & 3.1 & 2.9 & 2.06 & 0.21 & 3.28 & 2.3094 \\ 
J030631.7-000438.1 & 0.204 & 19.77 & 3.6 & 4.7 & 1.09 & 0.32 & 2.95 & 2.0806 \\ 
J030247.91-011203.5 & 0.194 & 19.36 & 4.6 & 3.0 & 4.55 & 0.38 & 3.27 & 1.8029 \\ 
J020618.92-011331.5 & 0.168 & 19.35 & 4.0 & 3.0 & 1.21 & 0.24 & 3.32 & 1.4210 \\ 
J030854.12-003651.5 & 0.149 & 18.85 & 4.8 & 4.8 & 2.33 & 0.43 & 3.24 & 1.4094 \\ 
J002458.18+004849.9 & 0.211 & 19.92 & 4.3 & 3.1 & 1.73 & 0.17 & 3.09 & 1.3265 \\ \hline
Green Valley $L_{*}$ Ellipticals\\
\hline
J204855.99-005506.1 & 0.097 & 18.19 & 3.9 & 4.1 & 0.72 & 0.00 & 3.14 & 10.0917 \\ 
J224321.35-001554.3 & 0.092 & 17.84 & 3.4 & 5.8 & 0.65 & 1.44 & 3.11 & 8.8189 \\ 
J211645.56-000017.7 & 0.092 & 17.88 & 5.4 & 3.1 & 2.08 & 1.64 & 2.95 & 4.9821 \\ 
J215116.94-011228.7 & 0.13 & 19.24 & 4.9 & 5.0 & 1.51 & 1.18 & 3.03 & 2.5047 \\ 
J000323.72+010547.3 & 0.099 & 17.88 & 4.0 & 4.9 & 0.91 & 0.38 & 3.23 & 2.0456 \\ 
J232319.22-002029.4 & 0.081 & 17.52 & 5.1 & 4.9 & 1.50 & 0.75 & 3.35 & 0.8173 \\ 
J212224.77+003436.4 & 0.113 & 18.35 & 3.3 & 3.3 & 1.52 & 0.57 & 3.50 & 0.5110 \\ 
J211933.3+010836 & 0.068 & 17.22 & 4.6 & 3.9 & 1.53 & 0.30 & 3.34 & 0.3707 \\ 
J234700.72-010124.1 & 0.134 & 19.01 & 4.4 & 4.3 & 1.45 & 0.00 & 3.41 & 0.0840 \\ 
J023042.42-003458.5 & 0.097 & 17.94 & 3.6 & 4.1 & 0.83 & 0.57 & 3.05 & \textbf{0.0039} \\ 
J211836.17-010831.8 & 0.061 & 16.99 & 6.0 & 6.1 & 2.29 & 0.00 & 2.94 & \textbf{0.0019} \\ \hline
Red $L_{*}$ Ellipticals \\
\hline
J030620.94-000344.5 & 0.112 & 18.94 & 3.2 & 2.8 & 1.71 & 1.29 & 2.99 & 2.7016 \\ 
J002539.99+000058 & 0.072 & 17.57 & 3.1 & 5.0 & 0.86 & 0.60 & 2.98 & 1.3446 \\ 
J022540.58+003728.4 & 0.073 & 17.53 & 5.6 & 5.2 & 2.58 & 0.34 & 3.06 & 0.5762 \\ 
J000055.43-010822.3 & 0.084 & 17.85 & 3.1 & 2.8 & 0.74 & 0.35 & 3.36 & 0.5186 \\ 
J213833.34-000058.5 & 0.061 & 17.10 & 4.3 & 3.6 & 1.99 & 1.01 & 3.49 & 0.3498 \\ 
J033251.29-002308.4 & 0.084 & 18.38 & 3.1 & 2.7 & 0.93 & 0.76 & 3.07 & 0.1906 \\ 
J024700.69+011018.6 & 0.067 & 17.27 & 5.1 & 5.3 & 1.15 & 0.50 & 3.37 & 0.1632 \\ 
J011532.22-001955.4 & 0.081 & 17.64 & 4.1 & 3.1 & 1.53 & 0.00 & 3.36 & 0.1295 \\ 
J223338.76-000119.7 & 0.087 & 18.12 & 4.4 & 5.2 & 1.87 & 0.00 & 3.14 & 0.0941 \\ 
J015654.84-003824.9 & 0.019 & 14.39 & 4.5 & 4.4 & 1.09 & 0.58 & 3.19 & 0.0387 \\ 
J222731.01-004011.2 & 0.059 & 17.26 & 3.6 & 3.2 & 1.23 & 0.38 & 2.91 & 0.0374 \\ 
J205830.81+010754.2 & 0.107 & 18.83 & 3.6 & 3.9 & 1.45 & 0.00 & 2.93 & \textbf{0.0047} \\ 
J221128.48-010439.4 & 0.090 & 18.50 & 3.2 & 2.5 & 0.73 & 0.00 & 3.36 & \textbf{0.0027} \\ 
J223505.75-010628.9 & 0.088 & 18.10 & 5.2 & 4.5 & 1.10 & 0.00 & 3.41 & \textbf{0.0024} \\ 
J020138.36-002425.3 & 0.081 & 17.86 & 3.4 & 5.1 & 0.79 & 0.20 & 3.17 & \textbf{0.0022} \\ 
J000239.72-004051.7 & 0.088 & 17.88 & 3.4 & 4.7 & 1.53 & 0.00 & 3.27 & \textbf{0.0021} \\ 
J213833.34-000058.5 & 0.066 & 17.23 & 3.5 & 2.5 & 1.38 & 0.00 & 3.35 & \textbf{0.0019} \\ 
J005120.44-005129.2 & 0.068 & 17.48 & 4.5 & 3.9 & 1.12 & 0.00 & 3.04 & \textbf{0.0018} \\ 
J002855.4-011416.8 & 0.085 & 18.03 & 4.4 & 2.8 & 2.03 & 0.00 & 3.39 & \textbf{0.0018} \\ 
J002914.95-000842.6 & 0.056 & 16.83 & 5.5 & 5.4 & 2.07 & 0.00 & 3.26 & \textbf{0.0015} \\ 
J220930.83-000914.6 & 0.057 & 17.02 & 5.1 & 4.6 & 1.62 & 0.00 & 3.19 & \textbf{0.0015} \\ 
J223211.1-002053.4 & 0.087 & 18.23 & 5.4 & 2.9 & 1.64 & 0.55 & 3.44 & \textbf{0.0015} \\ 
J005549.58-010453.8 & 0.045 & 16.43 & 3.1 & 3.8 & 0.62 & 0.00 & 3.29 & \textbf{0.0014} \\ 
J005507.25+005704 & 0.043 & 16.35 & 4.3 & 3.9 &  0.73 & 0.00 & 3.23 & \textbf{0.0012} \\ 
J005810.18-003733.8 & 0.045 & 16.37 & 4.1 & 3.5 & 0.46 & 0.00 & 3.33 & \textbf{0.0011} \\ 
J222814.39+003219.8 & 0.060 & 17.13 & 4.7 & 6.0 & 1.82 & 0.00 & 3.42 & \textbf{0.0011} \\ 
J232108.5+002430.2 & 0.031 & 15.49 & 4.3 & 3.1 & 1.05 & 0.00 & 3.41 & \textbf{0.0010} \\ 
J024526.61+005436.3 & 0.025 & 14.78 & 5.7 & 5.9 & 1.94 & 0.00 & 3.44 & \textbf{0.0007} \\ \hline
\end{tabular}
      \small
%        \item {\bf Note} - {Column (1) is the SDSS object Id, column (2); spectroscopic redshift, column (3); r band apparent magnitude, column (4); S$\acute{e}$rsic index, column (5); effective radius, column (6); color excess from balmer decrement method, column (7); stellar mass in units of 10$^{10}$M$_{\odot}$ and column (8); H$\alpha$ SFR (M$_{\odot}$/yr) within central 3 arcsec corrected for Balmer decrement. Values in bold denotes H$\alpha$ flux under 3 sigma rms.}
\caption{Properties of $L_{*}$ galaxies - Column (1) is the SDSS object Id, column (2); spectroscopic redshift, column (3); r band apparent magnitude, column (4) and (5); S$\acute{e}$rsic index in r and g bands, column (6); effective radius, column (7); color excess from balmer decrement method, column (8); stellar mass in units of 10$^{10}$M$_{\odot}$ and column (9); H$\alpha$ SFR (M$_{\odot}$/yr) within central 3 arcsec corrected for Balmer decrement. Values in bold denotes H$\alpha$ flux under 3 sigma rms.}        
\label{table:properties}
\end{center}
\end{table*}

%%%%%%%%%%%%%%%%%%%%%%%%%%%%%%%%%%%%%%%%%%%%%%%%%%%%%%%%%%%%%%%%%%%%%%%%%%%%%%%%%%%%%%%%%%%%%%%%%%%%%

%%%%%%%%%%%%%%%%%%%%%%%%%%%%%%%%%%%%%%%%%%%%%%%%%%%%%%%%%%%%%%%%%%%%%%%%%%%%%%%%%%%%%%%%%%%%%%
\subsection{$L_{*}$ elliptical galaxy sample}
\label{sec:lstar}

From our final galaxy sample, 1,109 galaxies are selected such that their masses lie in the $L_{*}$ mass range i.e., 2.9$\times$10$^{10}$ M$_{\odot}$ - 3.5$\times$10$^{10}$ M$_{\odot}$ \citep{Belletal2003,Kauffmannetal2003b}. This sample of $L_{*}$ galaxies further reduces to 960 galaxies as we discard galaxies with axis ratio (b/a) $>$ 0.5. The final sample of 960 $L_{*}$ galaxies are visually classified to identify their morphologies.
Of these, we found 766 Early type (E/S0), 84 spirals and 110 irregular/merging galaxies; a detailed morphological analysis of which will be presented in a forthcoming paper.

In order to characterize the nature of the light distribution in our Early-type $L_{*}$ galaxies, we have created the surface brightness profiles using IRAF \citep{Tody93} ellipse task on the SDSS r-band stamps. We model the 1D surface brightness profiles using a single S$\acute{e}$rsic function defined as:

\begin{equation}
  I(R) = I_e \exp\bigg\{-b_{n}\bigg[\bigg(\frac{R}{R_e}\bigg)^{\frac{1}{n}} -1\bigg]\bigg\},
  \label{eq:sersic}
 \end{equation}

\noindent where $R_e$ is the effective radius, $I_e$ is surface brightness at $R=R_e$; and $n$ denotes the S$\acute{e}$rsic index, a parameter measuring the concentration of light towards the central region. The parameter $b_n$ depends on n \citep{Graham2005}. A Moffat PSF with FWHM of 3 pixels corresponding to $\sim$ 1.1$^{\prime\prime}$ to 1.3$^{\prime\prime}$ in the r band has been used for convolution. Then the fitting was performed using Profiler \citep{2016PASA...33...62C} for all 766 Early-type galaxies. 

In addition, we have performed 2D fitting of the galaxies with single S$\acute{e}$rsic component using GALFIT \citep{Pengetal2002}. For this, we prepare, cut-out of galaxy images with size three times larger than the galaxy's petrosian radius. The size of a stamp is chosen to be large enough to contain sky for a good fit to the galaxy. Segmentation maps of the image stamps are created with SExtractor to mask neighbouring sources. Once we mask all the neighbouring sources, we perform a single S$\acute{e}$rsic fitting using PSF provided with the stacked data \citep{2016MNRAS.456.1359F} for the convolution. The parameters obtained from SExtractor are used as initial guess values for the fitting to proceed.
The best-fit parameters obtained from both 2D modelling (GALFIT) and 1D modelling (Profiler) are in good agreement with each other (see Appendix~\ref{sec:structure}). In Fig.~\ref{fig:compare}, a comparison of parameters (S$\acute{e}$rsic index and half light radius) obtained from 1D and 2D fit are shown. As 1D decomposition was done individually for each galaxy, we use the S$\acute{e}$rsic index, mean surface brightness and half-light radius from 1D modelling rather than 2D decomposition, which was performed in an automated way. 

% {\color{red} the following is not clear?? Why as expected?}\\
% As expected, since $L_{*}$ galaxies are in the low mass regime, the single component fitting is dominated by disc like light profiles i.e. S$\acute{e}$rsic index $\sim$ 1.

In the subsequent analysis, we select galaxies with S$\acute{e}$rsic index 3 $<$ n $<$ 6 as elliptical galaxies leading to a sample of 51 galaxies in this range. For the sake of comparison between filters, we have also performed 1D surface brightness profile fitting of the 51 $L_{*}$ ellipticals in SDSS g-band. The S$\acute{e}$rsic indices in the g-band (see Table~\ref{table:properties}) are slightly lower than those in the r-bands in some cases \citep[see][for a similar trend]{Vikaetal2012}. In Fig.~\ref{fig:sersic}, we show the histograms of the derived S$\acute{e}$rsic indices in g and r-bands with median values at 3.9 and 4.3 respectively. 

Based on the broadband photometric properties of 51 elliptical galaxies, they are further segregated into blue $L_{*}$ Ellipticals (BLEs), red $L_{*}$ Ellipticals (RLEs) and Green valley $L_{*}$ Ellipticals (GLEs) using the color magnitude diagrams (Fig.~\ref{fig:cmd}) with color cuts from \citet{Blanton2006}.
Blue galaxies are identified as

\begin{equation}
(g-r)_{blue} < 0.65 - 0.03\times(M_{r}+20)
\label{eq:blue}
\end{equation}

\noindent while red ellipticals as 

\begin{equation}
(g-r)_{red} > 0.80-0.03\times(M_{r}+20)
\label{eq:red}
\end{equation}

\noindent The intermediate ones are classified as galaxies in green valley. We found 12 BLEs, 28 RLEs and 11 GLEs. RGB image stamps, their corresponding surface brightness profiles along with S$\acute{e}$rsic fit and optical spectra of representative BLEs, GLEs and RLEs from the sample are shown in Fig.~\ref{fig:stamps}. The properties of all 51 $L_{*}$ ellipticals are discussed in the following sections and summarized in Table~\ref{table:properties}.

\section{Spectroscopic Analysis}
\label{sec:spec}
    We obtained the optical spectra of the 51 $L_{*}$ elliptical galaxies from SDSS DR12. The SDSS spectrum has a spectral coverage in the wavelength range of $3800 - 9200$ \AA~with a spectral resolving power of $R \sim 1800$. By modelling their spectra, we derive various spectral parameters viz. emission line fluxes, age of the stellar population and spectral indices such as D$_n$(4000) and H$\delta_{eqw}$. All the emission line fluxes used in the present study are derived by fitting Gaussian models to the observed emission profiles using the SPLOT task available in IRAF environment. Prior to this, we subtract the stellar continuum and underlying absorption features by modelling the spectra with the pPXF code \citep{CappellariEmsellem2004} which uses the MILES stellar libraries. Derived emission line fluxes are further corrected for reddening, due to both foreground \citep{2011ApJ...737..103S} and internal (host galaxy) dust extinction using the observed H$\alpha$ and H$\beta$ line flux ratio (i.e., Balmer decrement method) by assuming the theoretical ratio as 2.86 assuming Case-B recombination \citep{OsterbrockBochkarev89} with an electron temperature of $\sim$ 10$^{4}$K and electron density of 100 cm$^{-3}$. The stellar population age is directly derived from the pPXF fitting to the observed spectrum. In order to derive the spectral index D$_n$(4000), we use the classical definition \citep{Bruzual83}, which uses the ratio of average flux density in red $4050 - 4250$ and blue $3750 - 3950$ \AA~bands on either two sides of the 4000~\AA~break, after correcting for emission line contamination to these two bands. Similarly, for estimating H$\delta_{eqw}$ absorption index, we use the definition provided by \citet{WortheyOttaviani97}. According to this definition, the H$\delta_{eqw}$ index is estimated using the absorption line feature in the band $4083.50 - 4122.25$~\AA~and average pseudocontinuum in the bandpasses of $4041.60 - 4079.75$ (blue) and $4128.50 - 4161.00$~\AA~(red). Note that the H$\delta$ absorption feature is obtained after subtracting the contaminating H$\delta$ gas emission line.  

\section{Properties of $L_{*}$ ellipticals}
\label{sec:properties}

$L_{*}$ galaxies are found to display a wide variation in their structural and spectroscopic properties \citep{Kauffmanetal2003}. Fig.~\ref{fig:cmd} shows the color-magnitude diagram (CMD) and Kormendy relation for the sample of 51 $L_{*}$ ellipticals. The CMD is obtained from the absolute g-r integrated colors, which are corrected for dust extinction and k-correction, as discussed in Section~\ref{sec:methods}. All $L_{*}$ ellipticals are found to follow the Kormendy relation \citep{1977ApJ...217..406K} and their surface brightness profiles are well fitted with a single S$\acute{e}$rsic function, as mentioned earlier. The solid line in Fig.~\ref{fig:cmd} is from \citep{2009MNRAS.393.1531G} with slope 1.74, and the dashed lines are the 3$\sigma$ scatter of the distribution. A colour composite image, surface brightness profile and optical spectra of a BLE, GLE and RLE as an example are shown in Fig.~\ref{fig:stamps}. RLEs show spectra like typical early type galaxies (ETG) with prominent absorption lines and positive continuum slope. While the BLEs are seen to be dominated by emission lines, negative or flat continuum, similar to star-forming galaxies. In the following sections, we discuss the properties of these 51 $L_{*}$ ellipticals in detail.

%%%%%%%%%%%%%%%%%%%%%%%%%%%%%%%%%%%%%%%%%%%%%%%%%%%%%%%%%%%%%%%%%%%%%%%%%%%%%%%%%%%%%%%%%%%%%%%%%%%
\begin{figure*}
 \includegraphics[width=\textwidth]{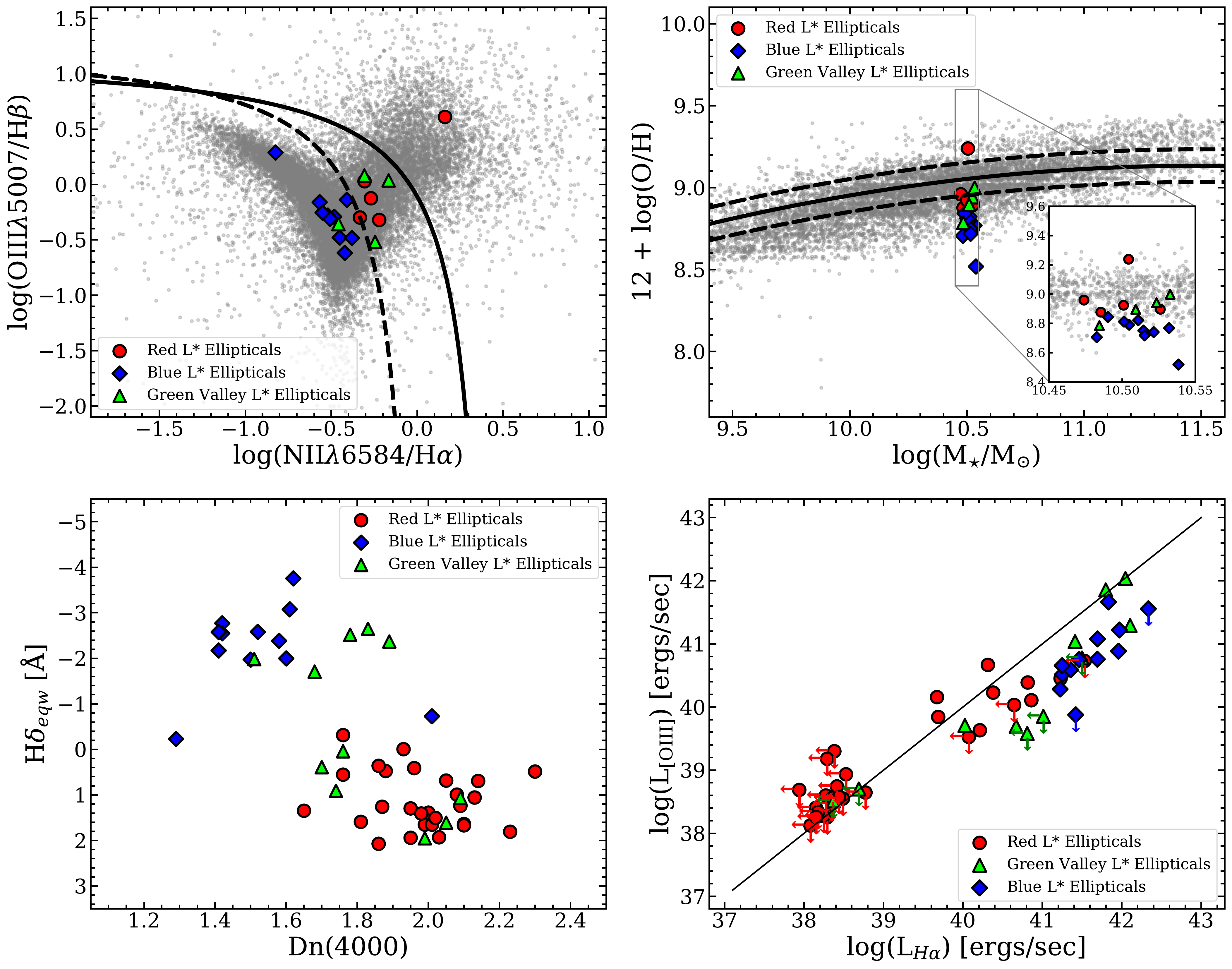}
 \caption{Emission line properties of $L_{*}$ ellipticals. Red circles represents red $L_{*}$ ellipticals, green triangles are green velley $L_{*}$ ellipticals and blue diamonds are blue $L_{*}$ ellipticals. Upper left panel BPT diagram - black solid line is the extreme star formation line \citep{Kewleyetal2001} while the dashed line is the pure star formation line \citep{Kauffmannetal2003}. Upper right, Mass-Metallicity Relation - Black solid line is from \citep{Tremontietal2004} and black dashed lines are 1$\sigma$ scatter (0.1 dex). Lower left, Dn(4000) vs H$\delta$ equivalent width. And lower right, OIII[5007] vs H$\alpha$ luminosity. Solid black is the x=y line, while the colored arrows indicate galaxies with line fluxes below three sigma rms.}
 \label{fig:bpt}
\end{figure*}

%%%%%%%%%%%%%%%%%%%%%%%%%%%%%%%%%%%%%%%%%%%%%%%%%%%%%%%%%%%%%%%%%%%%%%%%%%%%%%%%%%%%%%%%%%%%%
\begin{table*}
 
\begin{center}
%\caption{Line Fluxes of $L_{*}$ galaxies}
\begin{tabular}{lccccc}
\hline \hline
SDSS Obj ID & u-r & H$\alpha$ Lum. & [OIII] Lum. & [OII] Lum. & [NII6584] Lum.\\
   & & $\times$10$^{40}$ ergs/s & $\times10^{40}$ ergs/s & $\times10^{40}$ ergs/s & $\times10^{40}$ ergs/s \\
(1) & (2)  & (3) & (4) & (5) & (6) \\ \hline
Blue $L_{*}$ Ellipticals \\
\hline
J000919.89-010229.6 & 2.04 & 93.04 & 16.60 & 85.32 & 30.63 \\ 
J215022.22+001017.7 & 1.14 & 90.74 & 7.65 & 64.41 & 34.38 \\ 
J224259.05+004624.8 & 1.56 & 67.91 & 46.11 & 83.34 & 10.18 \\ 
J211651.79-010141.4 & 1.61 & 49.44 & 11.97 & 57.97 & 13.38 \\ 
J025652.34+004555.3 & 1.70 & 49.22 & 5.70 & 34.20 & 17.44 \\ 
J025302.59-011305.1 & 1.35 & 29.10 & 5.66 & 24.53 & 8.20 \\ 
J030247.91-011203.5 & 1.74 & 22.72 & 3.89 & 24.60 & 7.09 \\ 
J020618.92-011331.5 & 1.78 & 17.91 & 3.28 & 16.18 & 5.41 \\ 
J030854.12-003651.5 & 2.05 & 17.76 & 4.51 & 19.83 & 6.92 \\ 
J002458.18+004849.9 & 1.62 & 16.71 & 1.92 & 10.24 & 6.98 \\     \hline
Green $L_{*}$ Ellipticals\\
\hline
J023042.42-003458.5 & 1.98 & 127.16 & 19.50 & 133.43 & 44.28 \\ 
J000323.72+010547.3 & 2.02 & 25.77 & 10.86 & 26.23 & 12.70 \\ 
J211933.3+010836 & 2.49 & 4.67 & 0.49 & 3.65 & 2.66 \\ 
J234700.72-010124.1 & 2.74 & 1.06 & 0.51 & 1.36 & 0.72 \\  \hline
Red $L_{*}$ Ellipticals\\
\hline
J002539.99+000058 & 2.51 & 16.94 & 2.83 & 22.02 & 10.19 \\ 
J022540.58+003728.4 & 2.43 & 7.26 & 1.28 & 7.95 & 3.37 \\ 
J000055.43-010822.3 & 2.54 & 6.53 & 2.44 & 5.22 & 3.23 \\ 
J020138.36-002425.3 & 2.70 & 1.63 & 0.43 & 1.28 & 0.88 \\ 
J015654.84-003824.9 & 2.56 & 0.49 & 0.70 & 0.00 & 0.71 \\ \hline

\end{tabular}
      \small
 %       \item {\bf Note} - {H$\alpha$, OIII[5007] and OIII[3727+3729] line luminosities of the blue and green $L_{*}$ ellipticals having fluxes above three sigma rms, in units of 10$^{40}$ ergs/s corrected for balmer decrement.}
\caption{Line Fluxes of $L_{*}$ galaxies - H$\alpha$, OIII[5007] and OIII[3727+3729] line luminosities of the blue and green $L_{*}$ ellipticals having fluxes above three sigma rms, in units of 10$^{40}$ ergs/s corrected for balmer decrement.}        
\label{tab:flux}
\end{center}
\end{table*}

\subsection{Emission Line Properties and Star formation rates}
\label{sec:BPT}

We use the emission lines [O{\sc iii}]$\lambda$5007, H$\beta$, H$\alpha$ and [N{\sc ii}]$\lambda$6584 by modelling the SDSS spectra (see sec.~\ref{sec:spec}) to construct a BPT diagram \citep*{BaldwinPhillipsTerlevich81} for our $L_{*}$ ellipticals (Fig.~\ref{fig:bpt}). In the BPT diagram, we consider only those $L_{*}$ ellipticals in which emission line strengths of [O{\sc iii}]$\lambda$5007, H$\beta$, H$\alpha$ and [N{\sc ii}]$\lambda$6584  have signal-to-noise ratio (S/N) above 3.

Background galaxies (marked by grey filled circles) plotted in this diagram are from our parent sample from Stripe 82 (see sec.~\ref{sec:data}). All the BLEs (blue diamonds) fall in the star-forming region in the BPT diagram. They emit strong emission lines such as H$\alpha$ and [O{\sc iii}]$\lambda$5007 indicating active ongoing star formation. 11 out of 28 RLEs with measurable line strengths mostly occupy the composite region bounded by dashed \citep{Kauffmanetal2003} and solid line \citep{Kewleyetal2001}, while one of the RLEs lies in AGN-dominated region. The GLEs are located in the composite region, with one in the star-forming region. The presence of RLEs and GLEs in the composite region indicates optical AGN activity along with star formation activity in these galaxies. However, none of the $L_{*}$ ellipticals are detected either in  XMM-Newton\footnote{\url{https://heasarc.gsfc.nasa.gov/cgi-bin/W3Browse/w3browse.pl}} \citep{2012xmm} or Chandra\footnote{\url{https://cda.harvard.edu/chaser/}} X-ray \citep{2002Chandra} imaging catalog.
\par
The Star-formation rate (hereafter, SFR) for all the $L_{*}$ galaxies has been estimated using H$\alpha$ luminosity, which is sensitive to very recent star formation (on a timescale $\lesssim$10 Myr). We use the H$\alpha$ line luminosity from our modelling of the SDSS spectra corrected for internal dust extinction using Balmer decrement (discussed in section~\ref{sec:spec}) for SFR estimates using the \citep{Kennicutt98} relation that employs Kroupa \citep{Kroupa2001} initial mass function:
 
\begin{equation}
SFR (M_{\odot}yr^{-1}) = 7.93\times 10^{-42} L_{H\alpha} (ergs~sec^{-1}) 
\end{equation}

\noindent Column 8 in Table~\ref{table:properties} shows the H$\alpha$ SFRs in the central regions of our $L_{*}$ ellipticals. At the mean (median) redshift z=0.068(0.071) for the RLEs, the SDSS spectroscopic fibre corresponds to a circle of radius 1.95(2.03) kpc. The mean (median) effective radii $R_e$ of the RLEs are 1.35 (1.38) kpc. In other words, the SDSS fibre covers about $\sim 1.5 R_e$ of the RLEs. Similarly, for the BLEs being at z=0.172 (0.184), the mean (median) $R_e$= 2.3 (1.73) kpc and the SDSS fibre covers about twice their effective radii. The case with green-valley ellipticals is similar to blue ones. This clearly reveals that the measured SFRs, to a large extent, could be considered as the global SFR and not localized to the very nuclear region of the galaxy, as is the case for some local big elliptical galaxies. However, a more robust evidence for the extent of star formation would be narrow-band H$\alpha$ imaging \citep[e.g.,][]{Paswan2019} or integral field spectroscopy - none of which is currently available for the BLEs in our sample.

\begin{figure*}
 \includegraphics[width=0.8\textwidth]{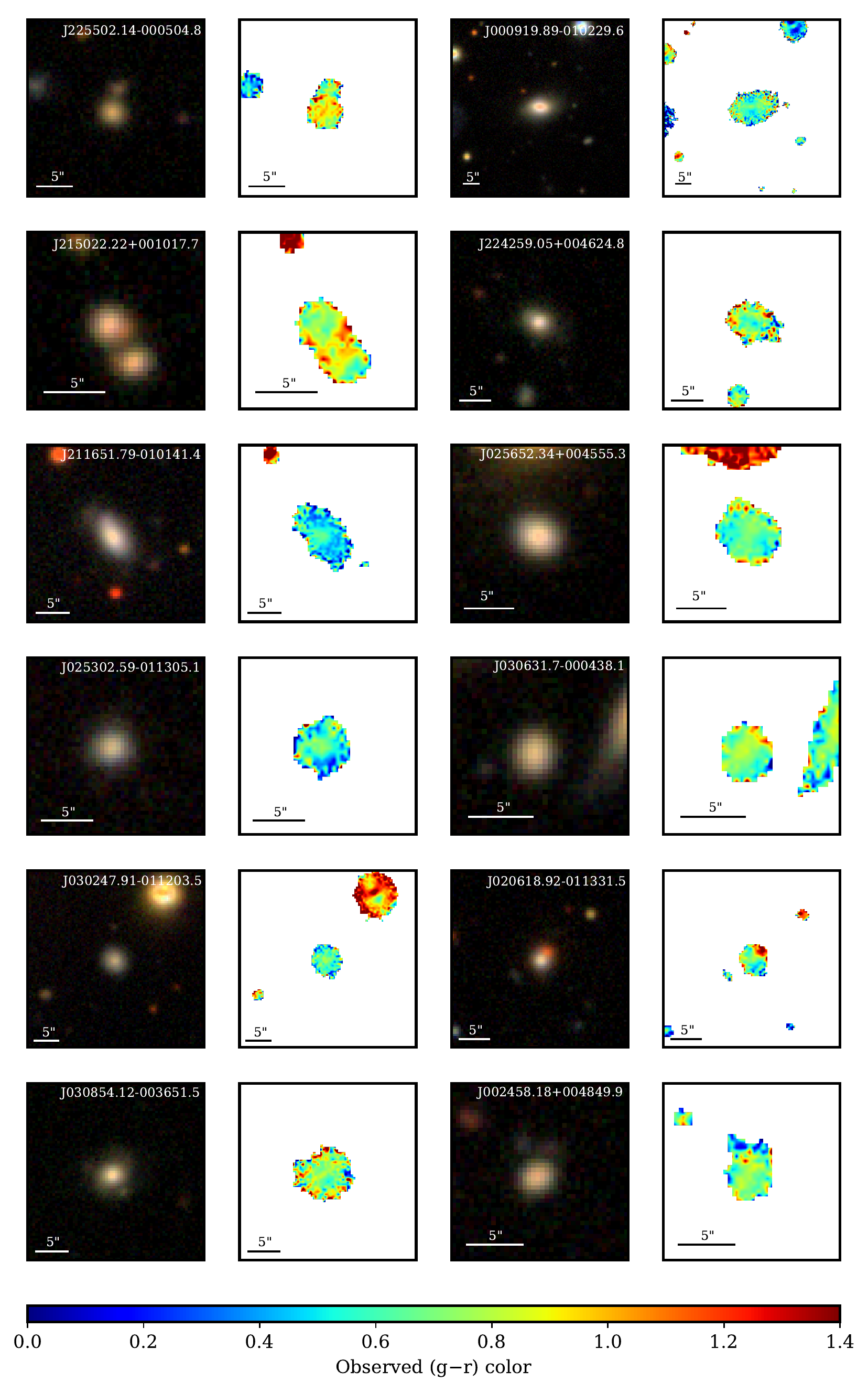}
 \caption{The gri color images and their corresponding (g-r) color maps for the 12 BLEs in the sample (see Table~\ref{table:properties} for their physical parameters). In the color map, sky pixels are masked. Horizontal bar on the bottom left measures 5$^{\prime\prime}$ scale on the sky.}
 \label{fig:cmap}
\end{figure*}

Normally, one would expect the young stars to be forming in a disk component. It might be possible that there are embedded disks in these galaxies, like the classic case of M87. If this is the case for the BLEs in our sample, the S$\acute{e}$rsic indices would be lowered than n=4. To that, we have re-analyzed the 51 ellipticals in the g-band. Table~\ref{table:properties} shows the g-band S$\acute{e}$rsic indices beside the r-band ones (see also histogram plot of the S$\acute{e}$rsic indices in the appendix~\ref{sec:structure}). For the BLEs and GLEs, there is no clear trend with filter change but the median $n$ is lowered by 0.4 from r to g-band. In other words, most of our BLEs have S$\acute{e}$rsic indices in the range of 3 - 6 in both r and g bands - making the BLEs even more intriguing. However, to investigate this further, observations in the Far Ultra-Violet (FUV) bands with a similar angular resolution as SDSS would be desirable. In the absence of narrow band H$\alpha$ and FUV imaging, the extended nature of star-formation in the blue L* ellipticals is derived from the optical (g-r) color maps. Fig.~\ref{fig:cmap} shows the gri color images and the corresponding (g-r) color maps on the same scale of the twelve BLEs in our sample. The color maps are created using the deep co-added sky subtracted images provided by \cite{ 2016MNRAS.456.1359F}. Since the images are sky subtracted, we mask the sky pixels using SExtractor segmentation maps to avoid negative flux values. At the L* mass range, according to \citep{Blanton2006} the (g-r) color cut for the blue cloud corresponds to a value of 0.68. Most BLEs from our sample show blue optical colors up to the galaxy outskirts, a strong evidence for the extended star formation in the BLEs.

The estimated SFR for the blue elliptical sample ranges from 1.32  to 17.13 M$_{\odot}$~yr$^{-1}$ with a mean SFR $\sim$ 4.60 M$_{\odot}$~yr$^{-1}$. In most of the RLEs, the SFRs are typical of the Early-type galaxies. The GLEs have on average low SFRs compared to BLEs, though with a few exceptions having strong emission lines and high H$\alpha$ SFR. 

\begin{figure*}
 \includegraphics[width=0.98\columnwidth]{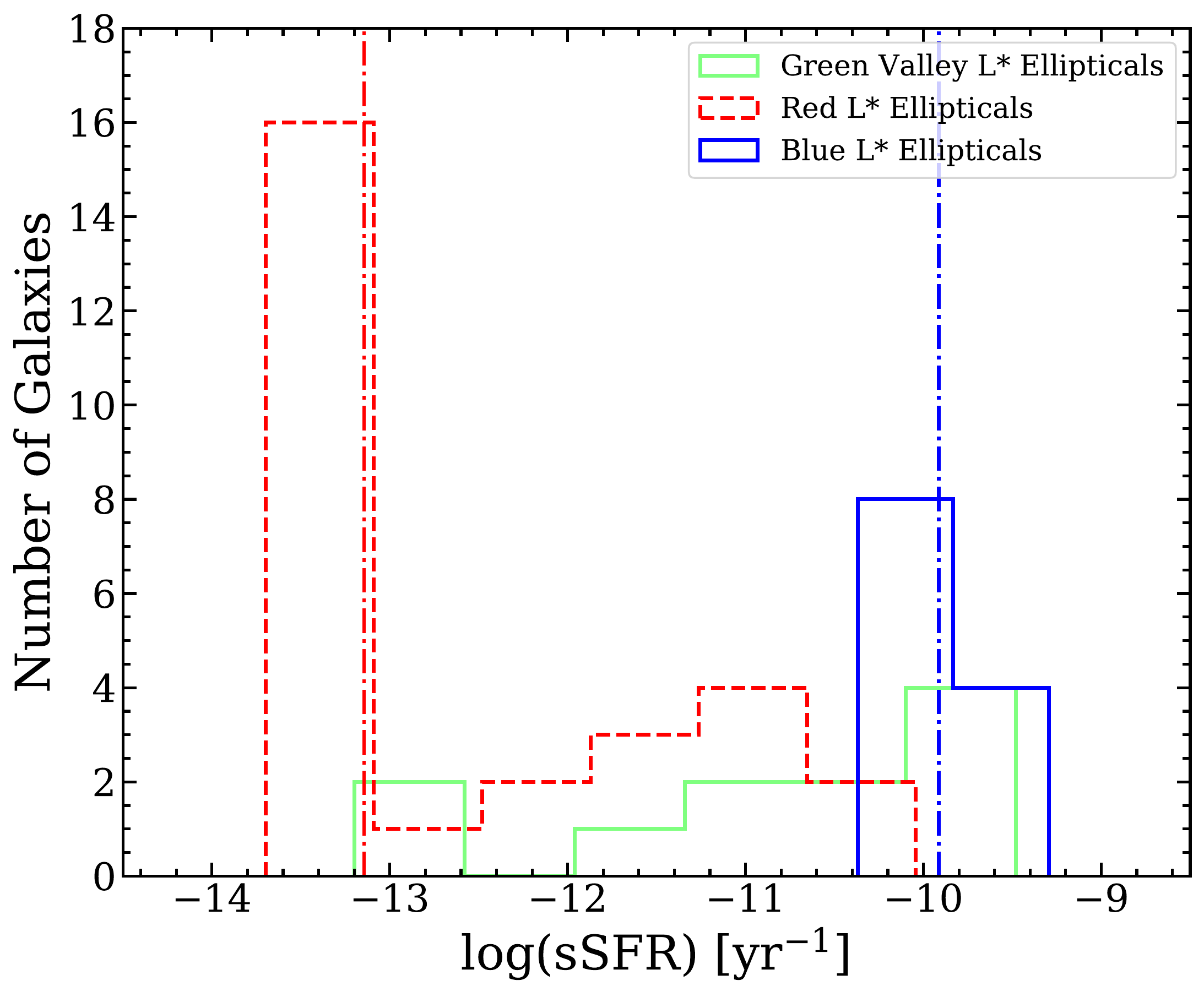}
 \includegraphics[width=\columnwidth]{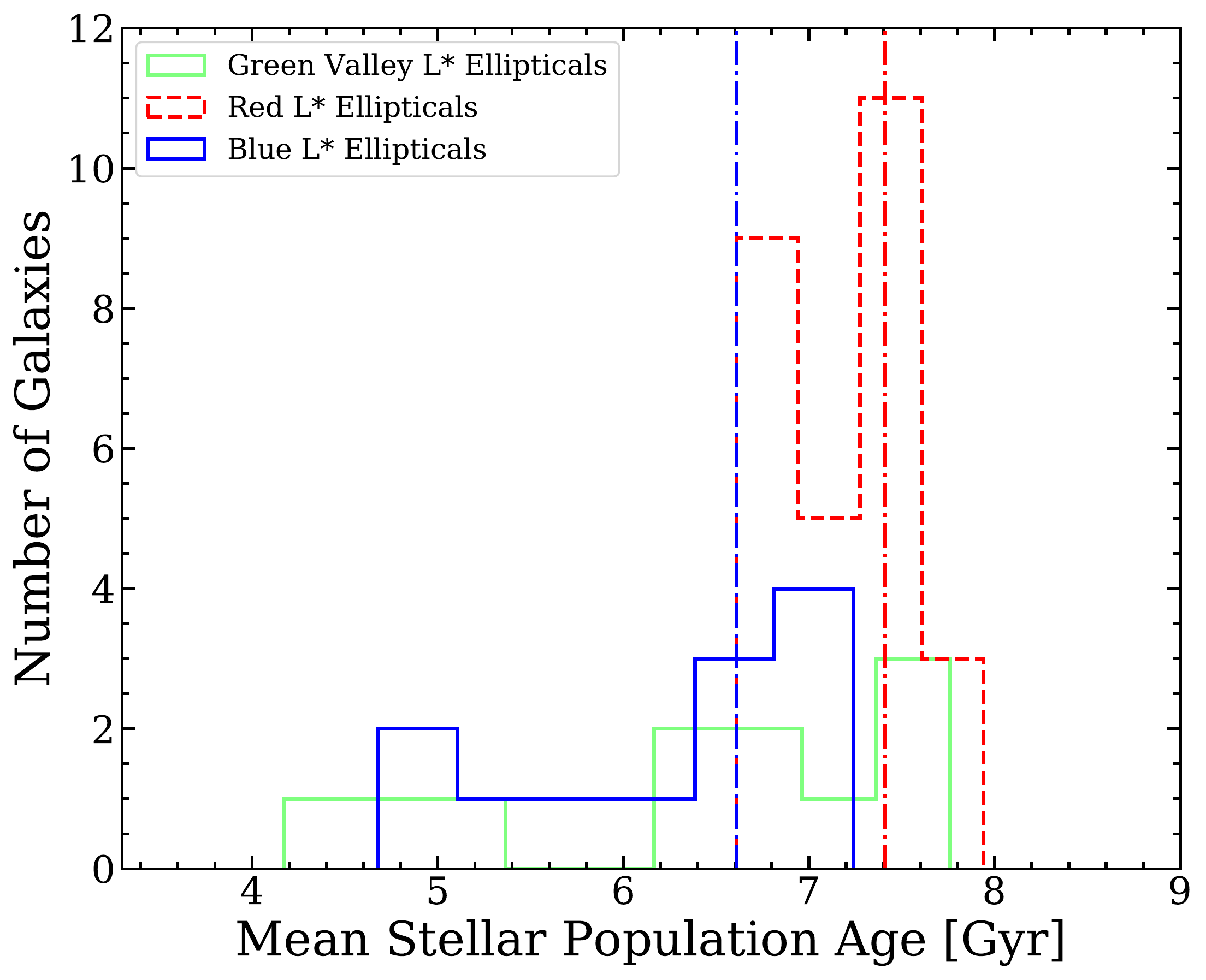}
 \caption{sSFR histogram- blue and red dashed dot lines represents the median values of blue and RLEs respectively. Right panel shows the mean stellar population age histogram for BLEs, GLEs and RLEs. The median being  6.61, 6.61 and 7.41 Gyr respectively.}
 \label{fig:ms}
\end{figure*}

Since the stellar masses of our $L_{*}$ ellipticals are around $3\times 10^{10}$~M$_{\odot}$, all the $L_{*}$ ellipticals in our sample occupy a thin vertical strip on the star-formation main sequence diagram \citep{Popessoetal2019}. At nearly constant stellar mass, the BLEs would evolve into the red ones following a vertical line on this diagram, essentially via the star-formation shut down. The median specific SFR (sSFR) for the BLEs is $\sim 10^{-9.9}$~yr$^{-1}$ (see Fig.~\ref{fig:ms}), placing them in the star-forming cloud while the median value of the sSFR of the RLEs is around $\sim 10^{-13.2}$~yr$^{-1}$  which would classify them as quenched population. Intriguingly, a few of the RLEs, apparently quenched, are still forming stars overlapping with the GLEs. In section~\ref{sec:quenching}, we discuss, in detail, the star-formation quenching in our sample of $L_{*}$ ellipticals.

%%%%%%%%%%%%%%%%%%%%%%%%%%%%%%%%%%%%%%%%%%%%%%%%%%%%%%%%%%%%%%%%%%%%%%%%%%%%%%%%%%%%%%%%%%%%%%%%%%%%%%%%%%%
 %%%%%%%%%%%%%%%%%%%%%%%%%%%%%%%%%%%%%%%%%%%%%%%%%%%%%%%%%%%%%

%%%%%%%%%%%%%%%%%%%%%%%%%%%%%%%%%%%%%%%%%%%%%%%%%%%%%%%%%%%%%%

%%%%%%%%%%%%%%%%%%%%%%%%%%%%%%%%%%%%%%%%%%%%%%%%%%%%%%%%%%%%%%%%
\begin{figure*}
 \includegraphics[width=0.95\textwidth]{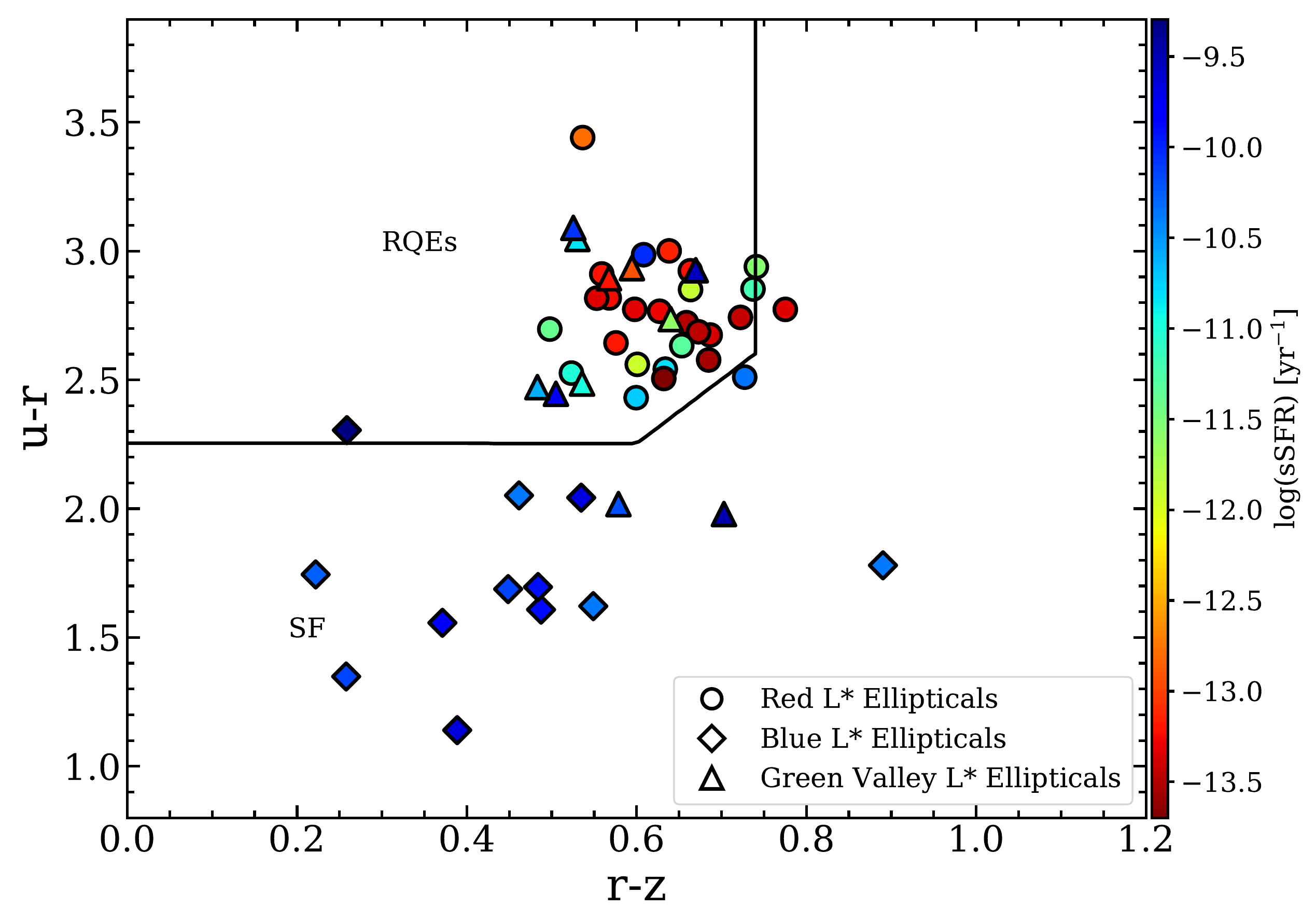}
 \caption{Recently Quenched Galaxies- Left panel. Black lines as defined by \citep{Holdenetal2012} separates the RQEs (inside boundary, u-r $\ge$ 2.25  and r-z $\le$ 0.74) from the star forming galaxies (outside boundary) based on the u-r and r-z color. Colour bar represents the sSFR.}
 %, whereas the standard deviation is 0.81, 1.15 and 0.40 Gyr.}
 \label{fig:urz}
\end{figure*}

 %#########################################################
 
\subsection{Gas Phase Metallicity and stellar population age}
\label{sec:metal}

The Mass-Metallicity Relation (MZR) reveals that metallicity generally correlates with galaxy stellar mass such that low-mass galaxies tend to have lower metallicities than their higher mass counterparts. MZR for the $L_{*}$ ellipticals is shown on the upper right panel of Fig.~\ref{fig:bpt}. Black solid line is from \citep{Tremontietal2004} and black dashed lines are 1$\sigma$ scatter (0.1 dex) lines. Metallicities for our $L_{*}$ ellipticals have been estimated using the NII[6584] and H$\alpha$ line fluxes following the calibration of \cite{Denicoloetal2002}:

\begin{equation}
12+log\bigg(\frac{O}{H}\bigg) = 9.12+0.73log(N2)
\end{equation}

\noindent Where N2 is log(\(\frac{\rm NII[6584]}{\rm H\alpha}\)). As expected, star-forming BLEs with high emission line fluxes have significantly low metallicities and are below the mass-metallicity relation with a mean metallicity value of 8.73. The RLEs and GLEs, on the other hand, have slightly higher or comparable metallicities to the BLEs (except one BLE with $\sim 8.5$ and RLE with $\sim 9.2$). An inset figure shows an enlarged view of the $L_{*}$ mass range and the metallicity distribution of the sample having a mean, median and standard deviation of 8.83, 8.82 and 0.08, respectively.
%{\color{red} starting abruptly.. without giving a perspective}

The stellar population age and star formation history can be reliably estimated using the 4000 \AA\ break: D$_n$(4000) and H${\delta}$ equivalent width \citep{Kauffmanetal2003}. Star-forming galaxies show weak D$_n$(4000) breaks, while passive stellar populations tend to have strong breaks. Similar to galaxy broadband color, D$_n$(4000) parameter separates these star-forming and passive galaxy populations. Whereas strong H${\delta}$ absorption lines are ubiquitous in galaxies that experienced a starburst activity $\sim 0.1-1$~Gyr ago. After which the galaxy luminosity in optical bands is dominated by late-type B to early-type F stars \citep*{Kauffmanetal2003, AngthopoFerrerasSilk2019}. Further, these two stellar indices are largely insensitive to the dust attenuation effects that complicate the interpretation of broadband colors. Hence, D$_n$(4000)-H$\delta_{eqw}$ plane turns out to be a powerful tool to study star formation history in galaxies. In Fig.~\ref{fig:bpt} (bottom left panel), D$_n$(4000) is plotted against H$\delta_{eqw}$ (+ve H$\delta_{eqw}$ values indicate H$\delta$ line in absorption, while -ve H$\delta_{eqw}$ values indicates H$\delta$ line emission) for our $L_{*}$ elliptical sample. BLEs form a separate group with low D$_n$(4000), while RLEs on the lower half with higher D$_n$(4000); indicate an older stellar population. \citet*{AngthopoFerrerasSilk2019} have found that older populations dominate at high stellar mass, and the star-forming systems dominate at low stellar mass. Galaxies with mass $\sim$3$\times$10$^{10}$M$_{\odot}$ are expected to lie in the lower region where the Dn(4000) is low, and galaxies have disc like morphology \citep{Kauffmannetal2003b}. Contrary to these results, we find that the RLEs, despite having the same mass range as that of BLEs, show strong 4000 \AA\ breaks and low H$\delta_{eqw}$. Hence, a diverse underlying stellar population. 

 We estimate the stellar population age of the galaxies by fitting the observed SDSS spectrum with pPXF code \citep{CappellariEmsellem2004} which uses the MILES stellar libraries. Fig.~\ref{fig:ms} shows the observed stellar population age histogram for all three groups of ellipticals in our sample. The median stellar population age for BLEs, GLEs and RLEs is 6.61, 6.61 and 7.41 Gyr, respectively, whereas the standard deviation is found to be 0.81, 1.15 and 0.40 Gyr, respectively. 

% The difference in the stellar population age of the three samples, especially the BLEs and RLEs. RLEs being dominated by older stellar population than the GLEs and BLEs as evident from Dn(4000) as well. 

%{\color{red} I think it is best to write the ages with 1sigma uncertainty instead of separately writing SD values.}

\section{Quenching in $L_{*}$ Ellipticals}
\label{sec:quenching}
Quenching plays an important role in the life-cycle of galaxies starting from the blue cloud (when galaxies are forming stars vigorously) to the red (and dead) sequence with little or no star formation. A number of physical mechanisms (mentioned in the introduction) have been put forward to understand the processes that lead to the star-formation shut down in a galaxy \citep{SommervilleDave2015}. Although there has been a large volume of work dedicated to understanding quenching, the exact physical mechanism, the time-scale of star-formation quenching and how it depends on various parameters of the host galaxy remain unknown to date. Equally important is to identify galaxies where quenching has just begun, i.e. the onset of quenching. Identifying galaxies in this transition phase will shed light on how galaxies transform\citep{Citroetal2017}. In this section, we discuss the properties of the $L_{*}$ ellipticals that are being recently quenched and identify a subset of $L_{*}$ ellipticals that are about to be quenched based on the physical state of their interstellar medium.

%\subsection{Recently Quenched Ellipticals}
\subsection{Ongoing quenching}
\label{sec:RQE}

 Recently Quenched Ellipticals (RQEs) are the post starburst (E+A, K+A) galaxies with strong Balmer absorption and little or no ongoing star formation \citep{Quinteroetal2004}. On the SDSS color-color (u-r) verses (r-z) plane, RQEs lie inside the boundary defined by colors,  u-r $\ge$ 2.25  and r-z $\le$ 0.74 (see Fig.~\ref{fig:urz}), while the star-forming galaxies occupy the region outside this boundary \citep{Holdenetal2012,McIntoshetal2014}. According to this color-cut, passive non star-forming galaxies are robustly separated from the blue star-forming ellipticals \cite{McIntoshetal2014}. All the RLEs, even the ones showing week H$\alpha$ emission in our sample are either inside or very close to the RQE boundary, whereas all the BLEs are found in the star-forming (SF) region (outside the RQE region). Of the 12 BLEs, only one is found to be crossing the u-r boundary (see Fig.~\ref{fig:urz}). The color separation works well as nearly all BLEs, star-forming as inferred based on their specific star-formation rate (sSFR), lie in the SF region and RLEs, being passive, inside the RQE region. GLEs are, on the other hand, found to be in both the RQE and SF regions. Two GLEs with [O{\sc{iii}}] emission are found to be outside the RQE boundary. Green-valley galaxies are thought to be the galaxies in transition \citep{Salim2014, Coendaetal2018, Nogueira-Cavalcanteetal2019, Angthopoetal2020}. The GLEs in our sample are no exception to this. Intriguingly, there are several RLEs and GLEs that lie inside the RQE boundary but are still star-forming. According to their sSFR values and presence of strong emission lines, they should be classified as star-forming ellipticals. In other words, although these $L_{*}$ ellipticals are recently quenched according to their broadband color selection, they should have been outside the RQE boundary. Six RLEs are found to be emitting H$\alpha$ with fluxes with $S/N > 3$ and are considered star-forming although they occupied the RQE region (see Fig.~\ref{fig:urz}). In other words, RQE criteria fails to classify these RLEs appropriately.   
 
 The median age of the underlying stellar population of the BLEs and GLEs are found to be $\sim 6.6$~Gyr, whereas the same for the RLEs is $\sim 7.4$~Gyr. The small difference in the stellar population age suggests that these $L_{*}$ elliptical galaxies have gone through a rapid transformation from blue to red via the green valley, and star formation has ceased not too long ago. Since we find some of these $L_{*}$ ellipticals in the transition phase, one might expect the star-formation quenching to be ongoing as well. The real question is whether one can predict which galaxies (especially among SF galaxies) will be quenched in the near future.  
 
 To further investigate the star-formation quenching mechanism and to identify which galaxies will be quenched, we closely examine the emission line properties in our sample of ellipticals using the SDSS fibre spectra following guidelines provided by \cite{Citroetal2017,Quaietal2018}. Based on the [{N\sc{ii}}] emission line, we found that the gas-phase metallicities of our $L_{*}$ ellipticals are similar around 8.8 with little scatter ($\sigma = 0.08$) except two galaxies (see the upper right panel of Fig.~\ref{fig:bpt}). The mean metallicity of GLEs and RLEs are 8.90 and 8.97, respectively. Being at similar metallicity, our sample is ideally suitable to investigate the state of star-formation quenching. According to \cite{Quaietal2018}, when star formation abruptly shuts down in a galaxy, the high ionization lines such as [{O\sc{iii}}] from O type stars are the first to dim down, followed by other comparatively low-ionization lines such as [{O\sc{ii}}] and then H$\alpha$. In other words, galaxies caught with these sequence of line strengths are the ones where quenching has begun, even if they are star-forming at present. We identify 10 BLEs, 4 GLEs and 5 RLEs (with $S/N > 3$ emission lines) for which [{O\sc{iii}}] line luminosity is lower than [{O\sc{ii}}] and H$\alpha$ (see Table~\ref{tab:flux}). This is also evident from the [{O\sc{iii}}]-H$\alpha$ luminosity diagram in Fig.~\ref{fig:bpt} showing a good fraction of the RLEs aligned with the star-forming BLEs and GLEs. 
 After carefully examining the emission line fluxes, we found 19 $L_{*}$ Es in our sample satisfy the following emission line sequence (Table~\ref{tab:flux}): 
 
 \begin{equation}
   [O III] < [O II] < H\alpha 
 \end{equation}

The above criterion based on emission lines is useful to identify galaxies in which the star-formation quenching might have just initiated. All the 19 galaxies (10 BLEs, 4 GLEs and 5 RLEs) are star-forming but are also the ones marked by the onset of quenching. In other words, these ellipticals could be termed as 'Soon to be Quenched Elliptical' (SQE). This sample of 19 SQEs would be extremely useful to understand how quenching proceeds in an elliptical galaxy. On the BPT diagram, the five RLEs fall in the composite region or AGN-dominated region. If the $L_{*}$ Es were quenched via starvation of cold gas and followed a close-box model \citep{Fabian2012, Ciconeetal2014, Fluetschetal2019, Trussleretal2020}, it would take about 3 Gyr. However, indications based on stellar population age estimate suggest that our RLEs quenched not too long ago. It remains to be unfolded what caused rather rapid quenching in these $L_{*}$ ellipticals. In the following, we probe their immediate environment to shed light on what might have caused such rapid quenching.      

\begin{figure}
 \includegraphics[width=\columnwidth]{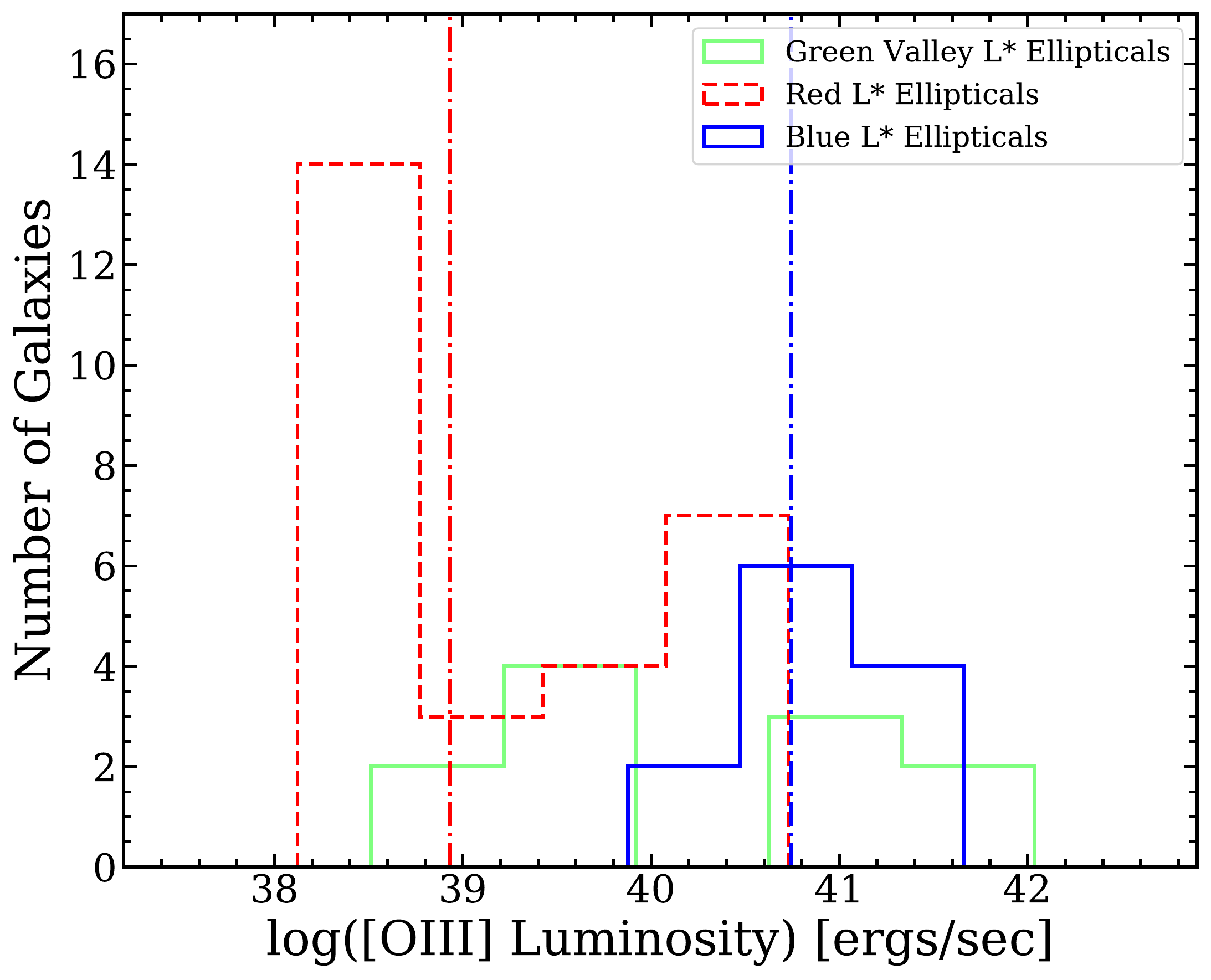}
 \caption{OIII[5007] luminosity histogram. blue and red dashed dot lines represents the median values of blue and red $L_{*}$ ellipticals respectively.}
 \label{fig:oiii}
\end{figure}

%%%%%%%%%%%%%%%%%%%%%%%%%%%%%%%%%%%%%%%%%%%%%%%%%%%%%%%%%%%%%%%%%%%%%%

\begin{figure*}
 \includegraphics[width=0.9\textwidth]{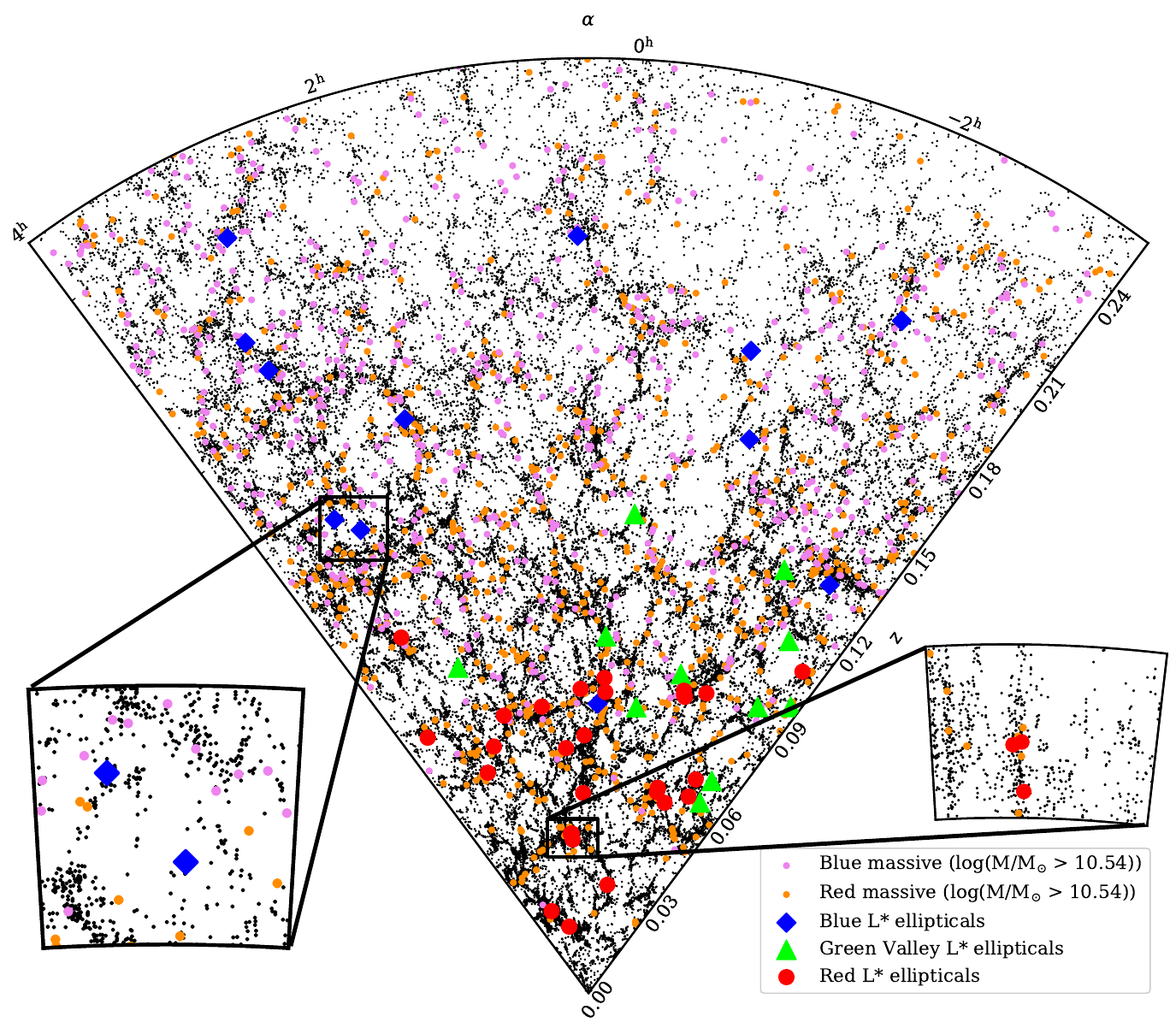}
 \caption{Large scale structure map. $L_{*}$ galaxies in our sample are shown along with $L_{*}$ galaxies from galaxy zoo. Inset figures show enlarged view of the environment around blue and red $L_{*}$ galaxies in our sample. Blue $L_{*}$ galaxies are in a comparatively low density regions while red $L_{*}$ ellipticals are in filaments/junctions.}
 \label{fig:lss}
\end{figure*}

\subsection{Impact of environment}
\label{sec:env}

To identify whether environment has played a role in the quenching of our $L_{*}$ ellipticals, we obtain a global picture of their neighbours on the cosmic web that might have been responsible for triggering galaxy-galaxy interactions \citep{Merritt1983}. Several methods have been used to estimate the number density of galaxies. One of the earliest ones was proposed by \cite{Fletcher1946}, while the most widely used is the friends-of-friends (FOF) algorithm \citep{HuchraGeller1982, MartinezSaar2002} which is based on linking length. Here, we follow a simple method to estimate the environment density around the $L_{*}$ ellipticals in the Stripe 82. We first calculate the local density around each $L_{*}$ galaxy. For this, we first compute the local volume at the redshift of a given galaxy using the following relation:

\begin{equation}
V_{*} = \Delta{RA} \times \Delta{Dec} \times (\frac{c {z_{*}}}{H_0}) \times \frac{c{\Delta z_{*}}}{H_0},
\label{eq:envdensity}
\end{equation}

\noindent in the above equation, $z_{*}$ is the redshift of the galaxy under consideration; $\Delta{RA}=\Delta{Dec}=0.6^{\circ}$ and $\Delta z_{*}=0.002$ or $z_{*}\pm 0.001$ around the galaxy. Then if there are $N_{*}$ galaxies within the volume $V_{*}$, the local density is simply $\rho_{*}=N_{*}/V_{*}$. In order to normalize the local galaxy density, we estimate the shell density at the redshift ($z_{*}$) of a chosen galaxy. Each shell is characterized by $110^{\circ}$ in RA and $\pm 2^{\circ}$ in Dec about the equator (since the data is from Stripe82 survey). Note that the redshift range of our $L_{*}$ Elliptical sample is $z_{*} \sim 0. - 0.24$. The thickness of each shell in the redshift direction is considered to be $\Delta z_{*} =0.004$ or $z_{*}\pm 0.002$ centered on the galaxy at $z_{*}$. Then the shell volume, $V_{*,shell}$, around each $L_{*}$ galaxy is estimated using the Eq.~\ref{eq:envdensity} but replacing $\Delta{RA}$ and $\Delta{Dec}$ by that of the shell. By counting the number of galaxies within each shell, say $N_{*,shell}$, we estimate the mean density of the shell as 

\begin{equation}
    \rho_{*,shell} = \frac{N_{*,shell}}{V_{*,shell}}
\end{equation}

\begin{figure}
    \includegraphics[width=\columnwidth]{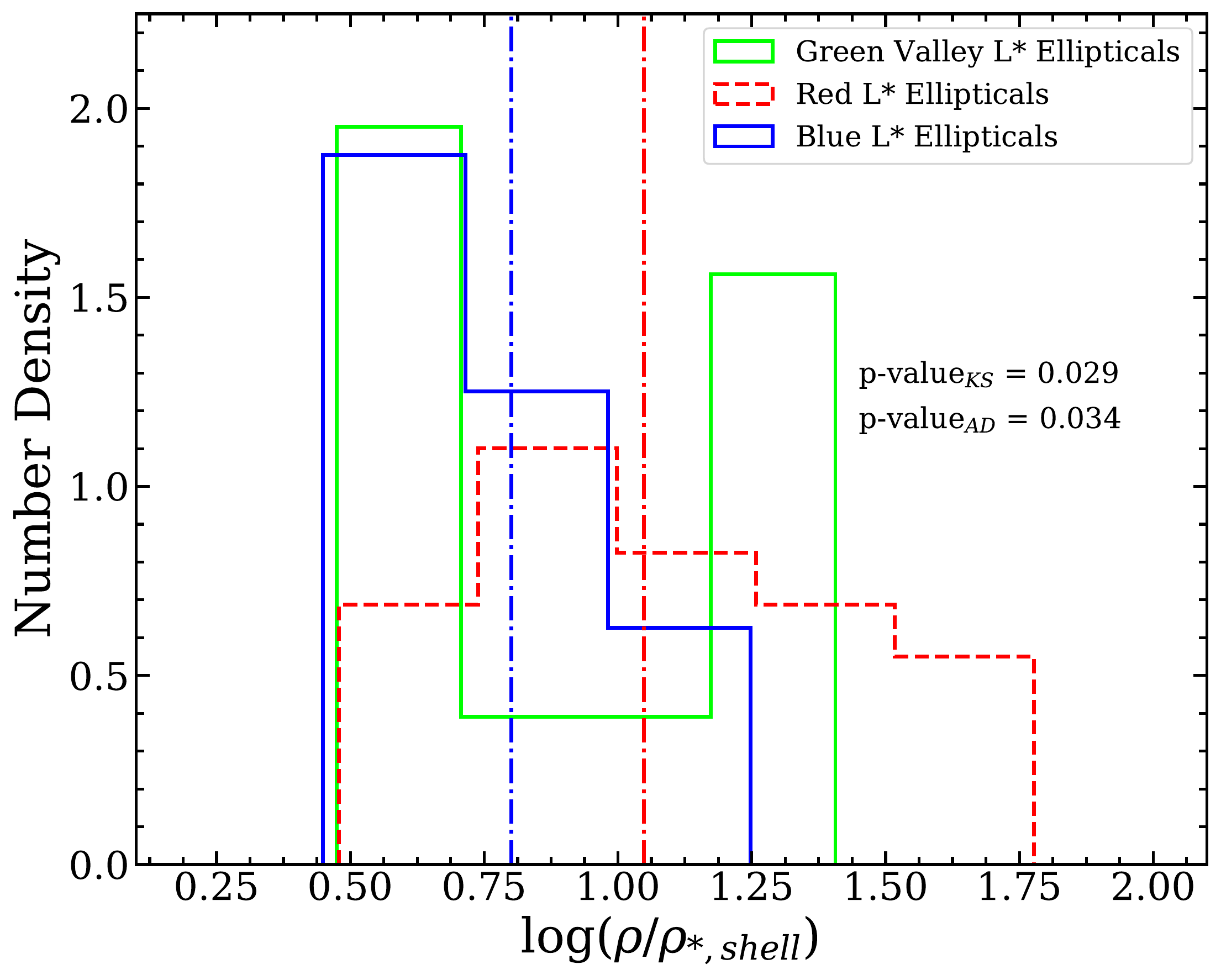}
  \caption{Environment density of galaxies - for the blue, green valley and red $L_{*}$ ellipticals. $\rho$ is the local density normalised by the shell density of galaxies. The KS- and AD-test show that the distributions of blue and red $L_{*}$ ellipticals are well separated at significance level = 0.05. Vertical blue and red dash dotted line are the medians of red and blue $L_{*}$ ellipticals respectively.}
  \label{fig:envdensity}
 \end{figure}

In Fig.~\ref{fig:envdensity}, we show the distribution of our $L_{*}$ Ellipticals (blue, green and red) as function of normalized environment density. From the histogram in the figure, it can be inferred that blue~$L_{*}$ ellipticals reside in comparatively low-density environment than the RLEs. On careful inspection of Fig.~\ref{fig:lss} and the zoom-in views, it appears that most of the BLEs are in the cosmic filament region, whereas the RLEs are preferably in the higher density region. Interestingly, the GLEs are found in both lower and higher density environment. These GLEs could be the missing link between the BLEs and RLEs. 
We perform the two tailed Kolmogorov-Smirnov test (KS-test) on the environment densities of the RLEs and BLEs to determine whether the two entities belong to the same distribution. The test gives a high D-statistics value (= 0.476) and a low p-value = 0.029, implying a moderately significant difference in the parent populations of the two samples at the significance level ($\alpha$) = 0.05. Further, a more sensitive Anderson-Darling test with p-value = 0.034 and high AD-statistics(= 2.387) reinforce our inference based on the KS-test. These tests, atleast, qualitatively confirms that BLEs preferably reside in a less dense environment than their red counterparts.  

It is likely that galaxies travel along the filament for a considerable fraction of their lifetime before emigrating to the denser region or cluster environment of the cosmic web. During this period, the galaxy interaction is expected to be infrequent and there might be plenty of cold gas available for nurturing these galaxies \citep{Keresetal2005, Hughesetal2013}. Our BLEs that are mostly in the filament are young and vigorously star-forming. As these BLEs travel along the filament and bump into a filament junction or cluster medium, they will be subject to ram pressure stripping of their cold gas reservoir \citep{Gunn-Gott1972} leading to star-formation shutdown, although it would depend on the density of the cluster environment; within a comparatively lesser dense medium, the process of gas removal would be less efficient and rapid star-formation shutdown might not occur in them \citep{Abadietal1999}. Most of the $L_{*}$ ellipticals have similar gas-phase metallicity $8.8 \pm 0.08$; with RLEs not deferring significantly from the rest  - implying strangulation \citep{PengMaiolinoCochrane2015} might not be the dominant process behind the BLEs transforming to RLEs. Besides, we know some RLEs (in the category of SQE) are still star-forming. Nevertheless, based on the stellar population age, we have clues that indicate RLEs did not quench a long ago. In other words, if BLEs were transformed to the RLEs, this timescale must have been in the last Gyr or so. In addition, most of our $L_{*}$ ellipticals (irrespective of their colors) are featureless, well-fitted by a single S$\acute{e}$rsic profile and have the same stellar mass range. 
Taking into consideration all these facts, it appears that our RLEs quenched through a process that is smooth enough to not alter their morphology, not change their stellar mass,  not to bring substantial signs of tidal interactions (see Appendix~\ref{sec:interacting}). Besides, some of the RLEs that are thought to be quenched based on their colors, are actually not; five such RLEs are in the category of SQE as ten BLEs and four GLEs are, sharing similar properties of their ISM. Since all these galaxies move along the cosmic web and eventually meet the junction or the cluster medium, the prevailing hot gas from the denser medium might have suffocated the BLEs. As a result, the cold gas in the BLEs heat up and eventually leads to the cessation of the star-formation without grossly changing any morphology or stellar mass. The end product of this process is probably our RLEs, while the GLEs remain as the intermediate cases.

\section{Discussion}
\label{sec:discuss}

%{\color{red} Please avoid repeating what has been said already, in the discussion. as much as possible}
%Can you add other ref than Martig et al? SD: Done

The transformation of galaxies from the blue cloud to the red sequence \citep{Conselice2014, Mishraetal2019} is one of the key phases in galaxy evolution. The question is, what causes this transformation and star formation quenching? There are several possible physical routes governing the quenching of the star-forming to passive galaxies, e.g. growth of bulges via secular evolution, stellar clump migration, violent disk instability transforming galaxies from late-type to early-type \citep{Lynden-Bell1979,Elmegreenetal2008,Sellwood2014,Ceverinoetal2015,Sachdevaetal2017, SahaCortesi2018,Martin-Navarroetal2018}, morphological quenching \citep{Martigetal2009,Wuytsetal2011, LillyCarollo2016, Luetal2021}, mergers \citep{RodriguezMonteroetal2019, Pawliketal2019}, environmental quenching \citep{BlantonMoustakas2009, Fillinghametal2018, Schaeferetal2019}. To pinpoint which of these processes contribute to the quenching and in what proportion is a difficult task. The motivation for studying a sample of ellipticals that undergo quenching is that it excludes the morphology change as the trigger for quenching, and thus may address the other reasons for the quenching.

It is widely accepted that galaxy evolution is strongly dependent on both stellar mass and environment \citep{Kauffmanetal2003, Brinchmann2004}. In this study, we consider only elliptical morphology in a very narrow mass range ($L_{*}$ mass range). This downsizes the sample dramatically but makes it an ideal sample to understand quenching in considerable detail. With this choice of sample, we are able to keep away the morphological transformation as well as those known to work efficiently in the high or low mass range, such as energetic feedback from active galactic nuclei or supernovae or energetic galactic winds \citep{Hopkinsetal2005, Veilleuxetal2005,Kavirajetal2007, Fabian2012,BoothSchaye2013,Kalinovaetal2021}. This leaves the halo-mass dependence of the $L_{*}$ ellipticals open for further investigation. The $L_{*}$ galaxies could be in clusters or in a field environment. 

\citet{McIntoshetal2014} in their study of recently quenched massive ellipticals (RQEs), considered a sample of galaxies at relatively lower redshift range up to z$\sim$ 0.08, and argued that the quenching in their sample has been primarily due to AGN feedback and mergers. Thanks to the deep coadded imaging from Stripe82 survey, we are able to identify $L_{*}$ ellipticals up to a redshift of z$\sim$0.25. $L_{*}$ elliptical galaxies in our sample do not show any sign of strong AGN feedback as indicated by the non-detection in X-ray emission through Chandra \citep{2002Chandra} and XMM-Newton \citep{2012xmm} imaging. However, optical AGN activity cannot be ruled out in GLEs and RLEs, as several of them lie in the composite region on the BPT diagram. Thus AGN activity could be contributing to the quenching processes in GLEs and RLEs. Nevertheless, it is interesting to point out that the reddening in these galaxies does not seem to be caused by AGN as the triggering of AGN is likely to be delayed compared to the onset of star formation, while galaxies are already reddened by consumption of gas by that time \citep{Hopkins2012, Yesufetal2014}. Furthermore, none of the BLEs shows any sign of AGN activity either in X-ray or the BPT diagram, implying that AGN is not the primary contributor to the quenching process in our L* ellipticals sample. 

\par
Galaxies subjected to interactions in dense environments like galaxy harassment \citep{Mooreetal96, Mooreetal98}, ram pressure stripping \citep{Gunn-Gott1972, Vollmeretal2001, Gavazzietal2013, Bosellietal2021} and galactic outflows, commonly show gas or stellar tails and asymmetries in morphology. Only a few of the $L_{*}$ ellipticals have close companions and show signs of interaction (See Appendix~\ref{sec:interacting}). The interacting features are not confined to a particular set of galaxies but are seen in all three sets of $L_{*}$ ellipticals. For instance, the three RLEs showing signs of interaction and companions have very low SFR or non-detection of H$\alpha$ line emission. One of these RLEs (J024526.61+005436.3) is interacting with a massive nearby galaxy. Four BLEs with interacting features have high SFR, while GLEs with such features have both high and low SFR. Since the quenching timescale in ram pressure stripped galaxies is short, $\lesssim$ 0.5-1 Gyr \citep{Boselli-Gavazzi2006, Yagietal2010, Gavazzietal2018, Liuetal2021}, it is tempting to assign ram pressure stripping as the primary cause of quenching in our sample of ellipticals, especially the RLEs because most of the RLEs are in the bracket of recently quenched ellipticals (RQE). However, some of these RLEs (with no signs of interaction) are still star-forming, having significant H$\alpha$ emission. So if we sum up the status of star-formation and associated signs of interaction, it seems unlikely that galaxy harassment and ram pressure stripping of cold gas are the primary causes of star-formation quenching in these ellipticals, especially the RLEs. But a further investigation based on cold gas measurement would be useful.

\par
Besides ram pressure stripping, galaxy harassment, and mergers, strangulation is another possibility considered to be causing quenching in galaxies \citep{PengMaiolinoCochrane2015}. However, strangulation brings a significant change in metallicity at the end of the quenching, albeit with a longer quenching timescale \citep[$\sim$3-4 Gyr;][]{Bosellietal2014}. Intriguingly, the metallicity of the $L_{*}$ ellipticals in our sample is $\sim 8.8$ with little scatter, in other words, metallicity values of the BLEs are not very different from RLEs in our sample. It is possible that strangulation is not the primary cause either. The useful input to these problems comes indeed from the SDSS fibre spectra of these galaxies. The RLEs that are already in the RQE region is characterized by the sudden disappearance of the [O{\sc {iii}}] emission. From the rest of the sample, we have selected a subsample of galaxies in which [O{\sc {iii}}] started going down - marking the onset of star-formation quenching \citep{Citroetal2017,Quaietal2018}. The triggering mechanism for this sudden quenching is yet to surface out. Further investigation, possibly with integral field spectroscopy which covers the outer parts of these galaxies.       

On the large scale structure map of galaxies in Stripe 82 region (see Fig~\ref{fig:lss}) up to a redshift of 0.25, BLEs are seen to be at relatively higher redshifts and in voids or sparse regions, whereas GLEs and RLEs are in denser regions (groups, filaments, clusters) and at lower redshifts. The Environment number density as presented in Fig.~\ref{fig:envdensity} also shows a similar trend. The redshift segregation is observed only in the case of $L_{*}$ ellipticals. This does not seem to be a selection effect. This discrepancy is not seen at higher and lower masses. In Fig.~\ref{fig:lss}, massive galaxies from the blue cloud and red sequence are plotted along with the $L_{*}$ ellipticals in which no such trend is seen. We are currently investigating whether such redshift distribution is generic.

Surprisingly, the blue ellipticals in the $L_{*}$ mass range are not found in any of the simulations. Most of the studies of star forming early type galaxies, consider a mass range higher than $L_{*}$ galaxies \citep{BirnboimDekel2003, Keresetal2005, DekelBirnboim2008, GaborDave2012, Nelsonetal2013, Feldmannetal2016}. While the lower mass counterparts or the sub $L_{*}$ galaxies have a disc-like or irregular morphologies \citep{Trujillo-Gomezetal2015}.  The local environment of galaxies does play a significant role in star formation quenching. In halos with masses $> 10^{11}$ M$_{\odot}$, hot gas and feedback processes like AGN heating dominates as indicated by cosmological hydrodynamical simulations \citep{BirnboimDekel2003, Keresetal2005, SijackiSpringel2006}. \citet{Angthopoetal2021} in their study of green valley galaxies in EAGLE and llustrisTNG with stellar mass log M/M$_{\odot} \lesssim$ 10 - 11, shows that green valley galaxies in EAGLE quench more rapidly and undergo later episodes of star formation, while the IllustrisTNG green valley galaxies exhibit more extended star formation history and quench at later cosmic times. They argue that the AGN feedback and the halo mass being the primary cause of the quenching; however, the study does not consider morphological classification.  We explored the IllustrisTNG100 simulation data at $z$ = 0 to search for BLEs. In the $L_{*}$ stellar mass range, we find 287 galaxies in TNG100 with primary$\_$flag = 1. We examined the visual morphology and the 2D S$\acute{e}$rsic fits provided by \citep{Rodriguez-Gomezetal2019} of the 287 galaxies to find that most of the galaxies in $L_{*}$ stellar mass range have disky or irregular morphology, while a few $L_{*}$ ellipticals found have redder colours and lie in the red sequence on the color-magnitude diagram. 

\section{Conclusions}
\label{sec:conclude}

Our primary conclusions are as follows:

\begin{itemize}
    \item We have selected a spectroscopic sample of 51 $L_{*}$ elliptical galaxies in the redshift range of $0 - 0.25$, chosen based on their visual morphology, S$\acute{e}$rsic index and Kormendy relation. In this sample, 12 are blue, star-forming having strong H$\alpha$, [{O\sc{iii}}] and [{O\sc{ii}}] emission lines. Four out of 11 green-valley and five out of 28 red ellipticals show similar emission lines as in the blue ellipticals.
    
    \item The SFR based on Halpha, using the SDSS fiber spectra, and the (g-r) color map reveals that the star-formation in BLEs is extended over the galaxy and not confined to their central region. The median specific SFR for the BLEs is $10^{-9.9}$~yr$^{-1}$, while for the RLEs, it is $10^{-13.2}$~yr$^{-1}$ - classifying the RLEs a quenched population.
    
    \item We find that most of the RLEs fall in the category of recently quenched ellipticals, i.e., the RQEs, while the BLEs are in an active star formation phase. A small fraction of the RLEs (5 out of 28) and the GLEs (4 out of 11) are in the star-forming phase, although they belong to the RQE population based on their broadband colors. 
   
    \item Based on the pPXF modelling of the SDSS spectra, we find that the difference in the median stellar population age of BLEs and RLEs is small, only ~0.8 Gyr. This small difference suggests that these $L_{*}$ elliptical galaxies have gone through a rapid star-formation quenching process.
    
    \item The RLEs and BLEs are clearly separated in the plane of $Dn(4000)$ versus H$\delta$ equivalent width (EW). The strong H$\delta$ absorption line in the BLEs suggests that they had a starburst in the last 0.1 - 1. Gyr.

    \item The BLEs are in relatively sparse environments, whereas RLEs are mostly found in denser regions such as the clusters at the junctions of filaments. Based on the spectroscopic analysis presented here, most of the RLEs have been quenched recently, with a few that are still forming stars while quenching has already started. The BLEs are the ones where star-formation quenching has just begun, while the green-valley ellipticals are intermediate between BLEs and RLEs.
    
    \item Although, galaxies go through a number of quenching processes and star forming episodes over their lifetime. In this study we are probing only the recent star formation episode in $L_{*}$ ellipticals, therefore it is possible that earlier star formation episodes may have quenched through other mechanisms.

\end{itemize}

%    We address the problem of environment on star formation quenching in $L_{*}$ elliptical galaxies and their evolution from blue cloud to red sequence. Despite of a very narrow mass range a wide range of spectroscopic properties have been observed in these galaxies.

%    \item Our study supports the claims that RQEs {\bf (RQEs being the ones which ceased star formation recently and show bluer optical, UV colors and fainter in IR magnitudes, whereas the red ellipticals are mostly red and dead with significant fraction of old stellar population)} have experienced a star burst activity.

%    \item {\bf The most probable way the $L_{*}$ blue ellipticals might have quenched star formation is through cutoff of cold gas while falling into the dense environment, clusters/filament junctions.} 

%   Although, galaxy evolution and star formation quenching is complicated subject and has several caveats. Our study of $L_{*}$ ellipticals gives a direction towards an evolutionary path of galaxies without structural transformation and mass acquisition.

%The correlation between colors and $Re$ of RQEs have also been reported by \cite{Wu2020}. 
%Star-Forming Main Sequence and the Mass-Metallicity Relation have been observed to hold in local universe as well as at redshift up to z=6 \citep{2004MNRAS.351.1151B, 2014ApJS..214...15S}.

\section*{Data Availability}
\addcontentsline{toc}{section}{Data Availability}
The SDSS co-added imaging data used in this paper is  publicly available at \url{http://www.iac.es/proyecto/stripe82}.

\section*{Acknowledgements}
% Entry for the table of contents, for this guide only
\addcontentsline{toc}{section}{Acknowledgements}
SD and KS acknowledges support from the Indian Space Research Organisation (ISRO) funding under project PAO/REF/CP167. SD and MBP acknowledges support from Department of Science and Technology (DST), New Delhi under the INSPIRE faculty  Scheme (sanctioned No: DST/INSPIRE/04/2015/000108). AD acknowledges support from the Israel Science Foundation grant 861/20. MBP gratefully acknowledges the support from the following funding schemes:  The Science and Engineering Research Board (SERB), New Delhi under the `SERB Research Scientists Scheme', ISRO under `AstroSat Data Utilization'  project.

%%%%%%%%%%%%%%%%%%%%%%%%%%%%%%%%%%%%%%%%%%%%%%%%%%



\bibliographystyle{mnras}
\bibliography{myref} % if your bibtex file is called example.bib

%%%%%%%%%%%%%%%%%%%%%%%%%%%%%%%%%%%%%%%%%%%%%%%%%%

%%%%%%%%%%%%%%%%% APPENDICES %%%%%%%%%%%%%%%%%%%%%

\appendix

%%%%%%%%%%%%%%%%%%%%%%%%%%%%%%%%%%%%%%%%%%%%%%%%%%%%%%%%%%%%%%%%%%%%%%%%%%%%%
\section{Structural parameters}
\label{sec:structure}

\begin{figure*}
 \includegraphics[width=0.95\columnwidth]{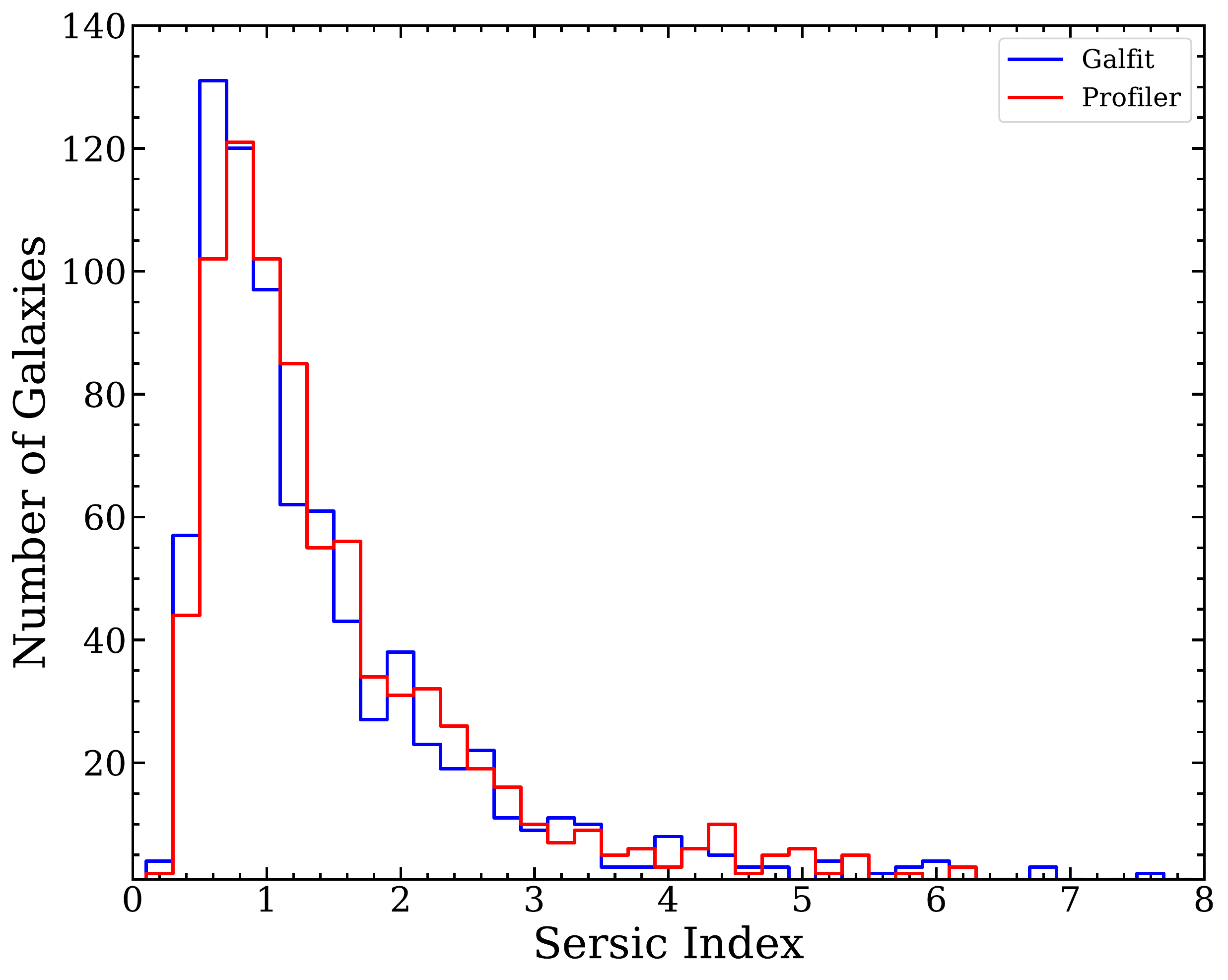}
 \includegraphics[width=0.95\columnwidth]{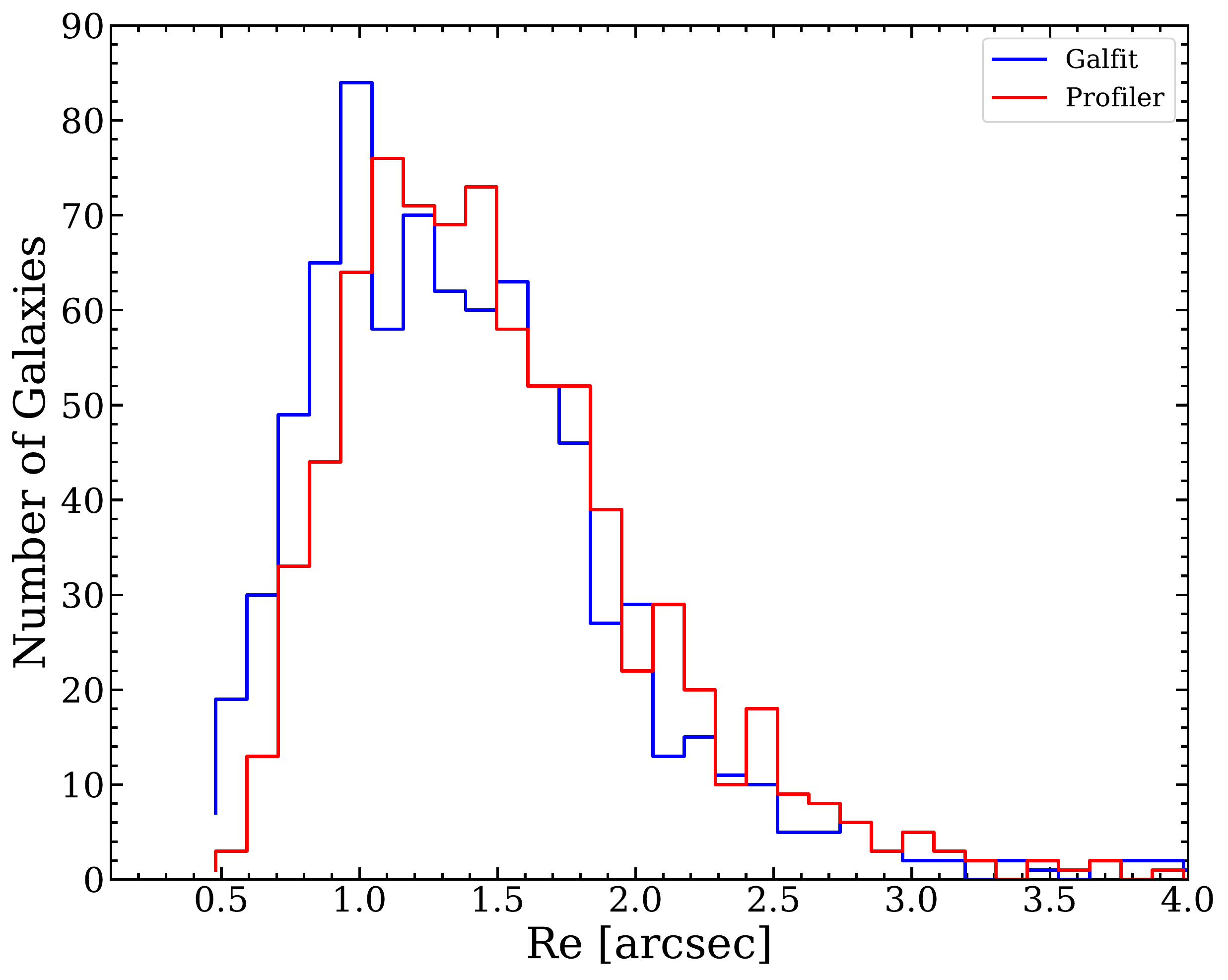}
 \caption{Comparison between the structural parameters (S$\acute{e}$rsic Index and Half light radius) obtained using 2D GALFIT fitting (blue) and 1D profiler fitting (red).}
 \label{fig:compare}
\end{figure*}

%%%%%%%%%%%%%%%%%%%%%%%%%%%%%%%%%%%%%%%%%%%%%%%%%%%%%%%%%%%%%%%%%%%%%%%%%%%%

\begin{figure}
    \includegraphics[width=0.95\columnwidth]{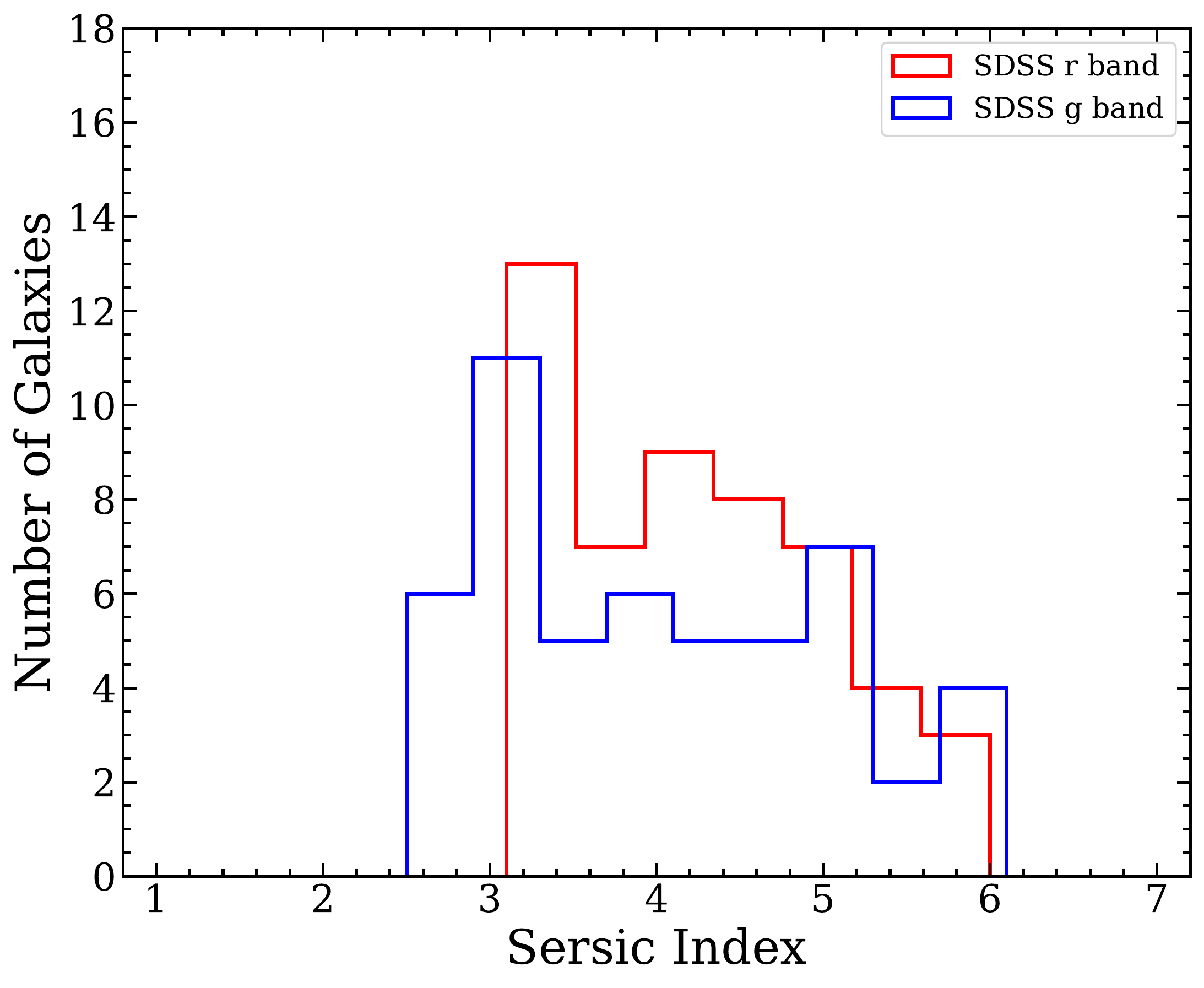}
    \caption{S$\acute{e}$rsic index derived from 1D profile fitting in SDSS r and g bands with median values 4.3 and 3.9 respectively.}
    \label{fig:sersic}
\end{figure}

%%%%%%%%%%%%%%%%%%%%%%%%%%%%%%%%%%%%%%%%%%%%%%%%%%%%%%%%%%%%%%%%%%%%%%%%%%%%

Fig.~\ref{fig:compare} shows the distribution of half light radius and S$\acute{e}$rsic index  obtained from GALFIT (2D) and Profiler (1D) structural fitting. 
We perform KS- and AD-test on both output parameters to examine if they belong to the same parent population. In case of S$\acute{e}$rsic indices, the resultant p-values corresponding to KS- and AD-test are 0.0128 and 0.009, respectively. On comparing half light radii, p-values for KS- and AD-test are 0. and 0.001, respectively. Although, at significance level = 0.05, the S$\acute{e}$rsic indices and half-light radii from 2D and 1D decomposition differ slightly from each other, the elliptical nature is evident in both the analysis.

Further, we show the S$\acute{e}$rsic index comparison of the 51 $L_{*}$ ellipticals in our sample, derived using the SDSS r and g band imaging data (Fig.~\ref{fig:sersic}). The S$\acute{e}$rsic index values show a good agreement in both the bands and most of the galaxies lie with the S$\acute{e}$rsic index values 3 < n < 6.

%%%%%%%%%%%%%%%%%%%%%%%%%%%%%%%%%%%%%%%%%%%%%%%%%%%%%%%%%%%%%%%%%%%%%%%%%%%%

 \begin{figure*}
    \includegraphics[width=\textwidth]{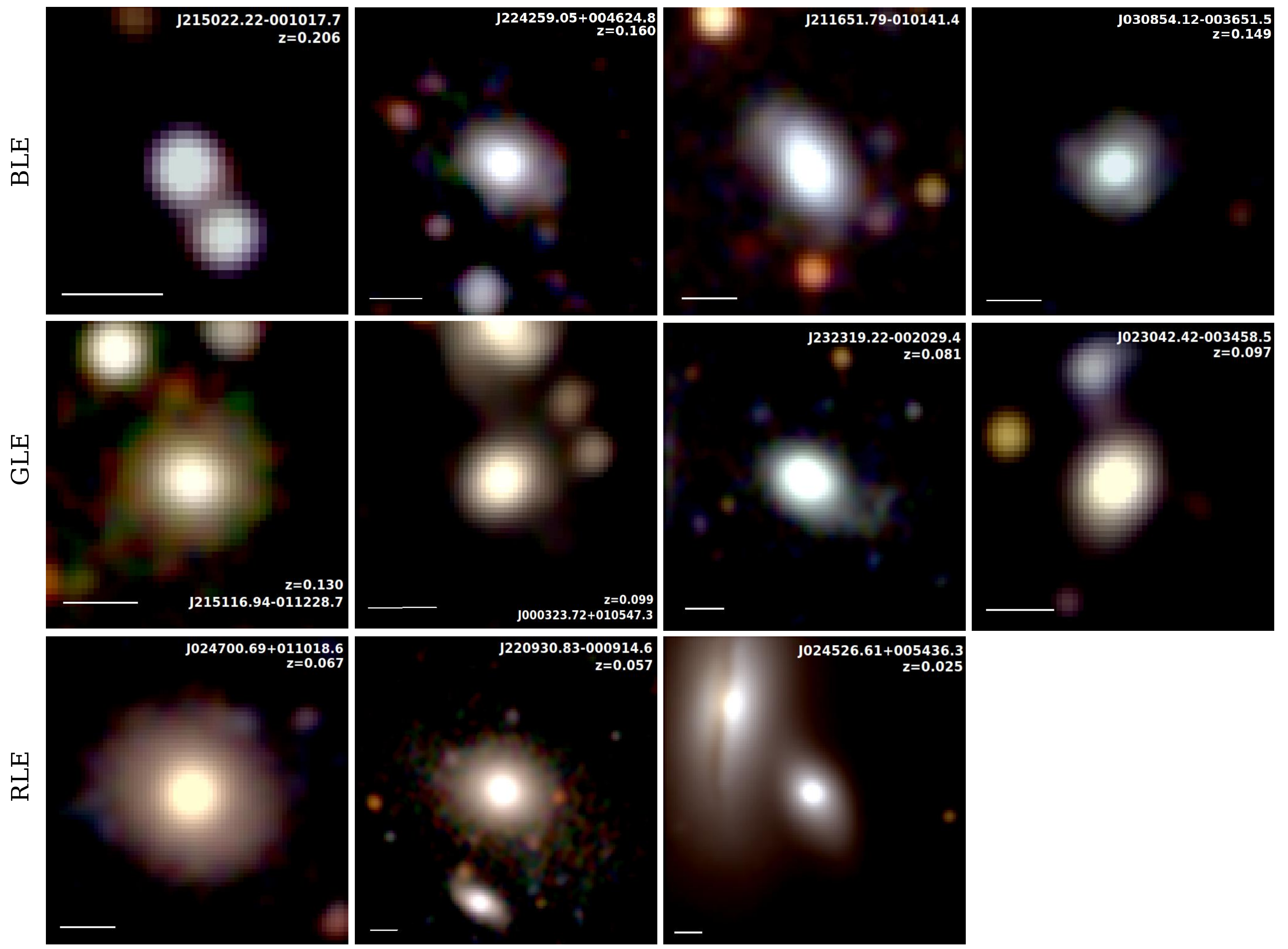}
  \caption{Galaxies from the $L_{*}$ sample showing interaction/merging features. Upper panel: blue $L_{*}$ ellipticals, middle: green valley $L_{*}$ ellipticals and lower panel: red $L_{*}$ ellipticals are shown respectively. horizontal bars on the left corner show 5$^{\prime\prime}$ scale on the image.}
  \label{fig:interacting}
 \end{figure*}

\section{Signs of Interaction in L* Ellipticals}
\label{sec:interacting}

Mergers or interaction with companion galaxies being one of the major mechanisms contributing in the process of quenching. While, gas accretion fueling the star formation activity, enhances the SFR \citep{George2017}. For the sample of 51 $L_{*}$ elliptical galaxies, we try to see signs of interactions/mergers. We see a few galaxies having faint tidal features, clumps around the galaxy or close companions. The images of those $L_{*}$ ellipticals are shown in Fig.~\ref{fig:interacting}. The details of the features or nearby companions of galaxies are as follows.\\

\noindent{\bf Blue $L_{*}$ ellipticals} (top panel of Fig.~\ref{fig:interacting})\\

\noindent{\bf\emph{J215022.22-001017.7}}\\
This BLE has an H$\alpha$ SFR of 7.2 M$\odot$ yr$^{-1}$. The companion galaxy seen at the lower right, has a redshift 0.213 (photo-$z$) while the BLE at the centre is at a redshift of 0.206 (spec-$z$). Considering the uncertainties on photo-$z$, both are probably at similar redshifts and interacting.\\

\noindent{\bf\emph{J224259.05+004624.8}}\\
This BLE has an H$\alpha$ SFR of 5.4 M$\odot$ yr$^{-1}$ and shows some extended asymmetric feature.\\

\noindent{\bf\emph{J211651.79-010141.4}}\\
This galaxy has an H$\alpha$ SFR of 3.9 M$\odot$ yr$^{-1}$, showing similar extended feature along the major axis. Other bright objects in the image are foreground stars.\\

\noindent{\bf\emph{J030854.12-003651.5}}\\
The $L_{*}$ galaxy has an H$\alpha$ SFR of 1.4 M$\odot$ yr$^{-1}$. There are two blob-like features very close to the galaxy. Most likely are associated with the galaxy, the redshift of the blobs could not be confirmed.\\

\noindent{\bf Green Valley $L_{*}$ ellipticals} (middle panel of Fig.~\ref{fig:interacting})\\

\noindent{\bf\emph{J215116.94-011228.7}}\\
The GLE has an H$\alpha$ SFR of 2.5 M$\odot$ yr$^{-1}$. An extended plume like feature is seen associated with the galaxy. The object on top left is a foreground star while on top right is a galaxy at redshift of 0.4 (spec-$z$).\\

\noindent{\bf\emph{J000323.72+010547.3}}\\
The $L_{*}$ galaxy has an H$\alpha$ SFR of 2.0 M$\odot$ yr$^{-1}$. The $L_{*}$ elliptical at the centre is at a redshift of 0.099, is interacting with the galaxy on top at the same redshift. Other two objects on the right are galaxies at redshift $\sim$0.4 (photo-$z$).\\

\noindent{\bf\emph{J232319.22-002029.4}}\\
The $L_{*}$ galaxy has an H$\alpha$ SFR of 0.8 M$\odot$ yr$^{-1}$. Extended tail on the lower left.\\

\noindent{\bf\emph{J023042.42-003458.5}}\\
The GLE does not have a significant H$\alpha$ line emission. It is at a redshift of 0.097, the galaxy on top is at a photo-$z$ of 0.099. The pair seems interacting. The galaxy on left is at photo-$z$=0.4 while the object at bottom is a star.\\

\noindent{\bf Red $L_{*}$ ellipticals} (bottom panel of Fig.~\ref{fig:interacting})\\

\noindent{\bf\emph{J024700.69+011018.6}}\\
This RLE has an H$\alpha$ SFR of 0.1 M$\odot$ yr$^{-1}$. Faint clumps are seen close to the galaxy.\\

\noindent{\bf\emph{J220930.83-000914.6}}\\
The $L_{*}$ galaxy does not have H$\alpha$ emission. Clumps are seen near the galaxy and extended light in the direction of position angle, especially on the lower side of the galaxy. Also the galaxy at the bottom is at the same redshift as the $L_{*}$ elliptical at z$\sim$0.057.\\

\noindent{\bf\emph{J024526.61+005436.3}}\\
The $L_{*}$ galaxy does not have H$\alpha$ emission, it is interacting with massive nearby galaxy at the same redshift $\sim$0.025.

\bsp	% typesetting comment
\label{lastpage}
\end{document}